# Tracing attosecond electron emission from a nanometric metal tip

Philip Dienstbier[1], Lennart Seiffert[2,*], Timo Paschen[1,3,*], Andreas Liehl[4], Alfred Leitenstorfer[4], Thomas Fennel[2,5,6], and Peter Hommelhoff[1]

[1]*Department of Physics, Friedrich-Alexander-Universität Erlangen-Nürnberg (FAU), Staudtstraße 1, D-91058 Erlangen, Germany, EU*
[2]*Institute for Physics, Rostock University, Albert-Einstein-Straße 23–24, D-18059 Rostock, Germany, EU*
[3]*Now with: Korrelative Mikroskopie und Materialdaten, Fraunhofer-Institut für Keramische Technologien und Systeme IKTS, Äußere Nürnberger Straße 62, D-91301 Forchheim, Germany, EU*
[4]*Department of Physics and Center for Applied Photonics, University of Konstanz, D-78457 Konstanz, Germany, EU*
[5]*Max Born Institute, Max-Born-Straße 2A, D-12489 Berlin, Germany, EU*
[6]*Department of Life, Light and Matter, University of Rostock, Albert-Einstein-Straße 25, D-18059 Rostock, Germany, EU*

**Solids exposed to intense electric fields release electrons through tunnelling. This fundamental quantum process lies at the heart of various applications, ranging from high brightness electron sources in DC operation[1,2] to petahertz vacuum electronics in laser-driven operation[3-8]. In the latter process, the electron wavepacket undergoes semiclassical dynamics[9,10] in the strong oscillating laser field, similar to strong-field and attosecond physics in the gas phase[11,12]. There, the sub-cycle electron dynamics has been determined with a stunning precision of tens of attoseconds[13-15], but at solids the quantum dynamics including the emission time window has so far not been measured. Here we show that two-colour modulation spectroscopy of backscattering electrons[16] uncovers the sub-optical-cycle strong-field emission dynamics from nanostructures, with attosecond precision. In our experiment, photoelectron spectra of electrons emitted from a sharp metallic tip are measured as function of the relative phase between the two colours. Projecting the solution of the time-dependent Schrödinger equation onto classical trajectories relates phase-dependent signatures in the spectra to the emission dynamics and yield an emission duration of $710 \pm 30$ attoseconds by matching the quantum model to the experiment. Our results open the door to the quantitative timing and precise active control of strong-field photoemission in solid state and other systems and have direct ramifications for diverse fields such as ultrafast electron sources[17], quantum degeneracy studies and sub-Poissonian electron beams[18-21], nanoplasmonics[22] and petahertz electronics[23].**

Lightwave electronics in the form of petahertz vacuum nanoelectronics is almost a reality: The waveform-sensitive results at freestanding metal nanostructures[9,10,24] indicated that these systems may provide switching frequencies and sampling bandwidths in the petahertz range. Combining the ultrafast response with nanostructuring techniques led to ultrafast light-driven metal nanostructures on substrates[3-8], which may enable lightwave electronic devices for signal-processing at optical clock rates. It has been shown that such devices allow on-chip control of electron transport within one optical cycle[3,4,6,7], are sensitive to the carrier-envelope phase[3,4,6,7], can sample optical waveforms[8] and down-convert optical waveforms to electronic signals[5]. Recently, the duration of the backscattered electron wavepacket could be measured[25]. Yet, the quantitative timing and duration of the sub-cycle electron emission from nanostructures is unknown, albeit it is fundamentally important for quantifying the maximum supported bandwidth, for example. So far, only qualitative insights exist based on, for example, the loss of photon order contrast near the cut-off in above threshold photoemission, hinting to a single temporal emission window. Notably, the width of this window could only be estimated so far[9].

In the semiclassical picture, strong-field photoemission begins with the release of an electron from the target via tunnelling followed by its propagation under the influence of the driving laser field. Upon re-encounter with the parent target, the electron can recombine under the emission of high harmonic (HH) and attosecond pulses[11,12], or can rescatter and escape where elastic backscattering typically yields maximal electron energies[16]. The analysis of the HH radiation as well as the rescattered electrons enables the characterization of both the electron emission process as well as the subsequent field-driven electron wavepacket dynamics.

---

[*]Both authors contributed equally.

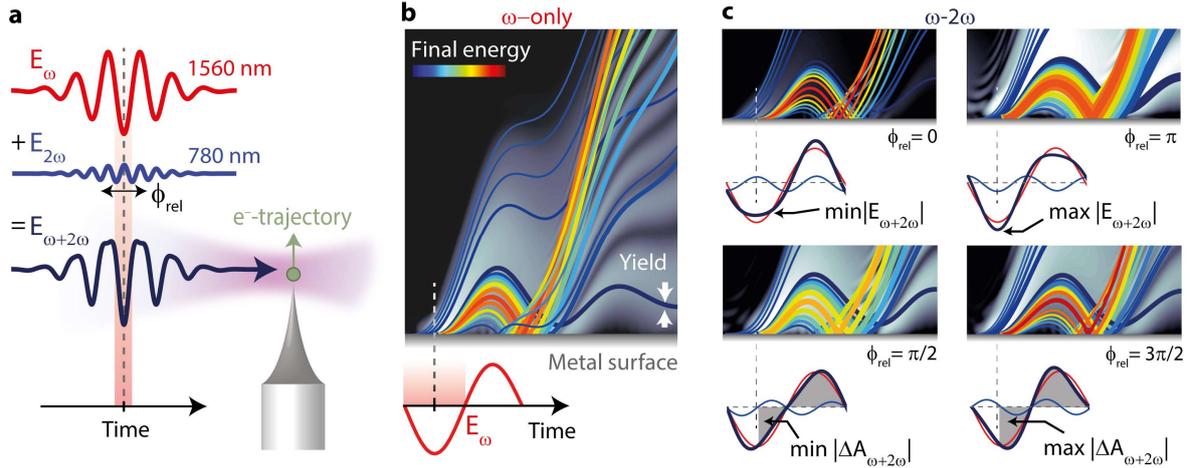

**Figure 1 | Schematic representation of the experiment. a,** A two-colour field ($E_{\omega+2\omega}$, dark blue) comprised of few-cycle pulses at 1560 nm ($E_\omega$, red) and 780 nm ($E_{2\omega}$, blue) with variable relative phase $\phi_{rel}$ emits electrons form the apex of a tungsten needle tip. **b,** Strong-field electron emission during the central half-cycle of the fundamental field (red shaded area) with successive electron propagation in the laser field and, for part of the trajectories, elastic backscattering at the surface. Curves represent classical trajectories with final kinetic energies and rates (yield per birth time duration) encoded by colour and linewidth, respectively. The background grey scale shows the electron density obtained from the TDSE simulation (white: many electrons, black: none, logarithmic scale), which is projected on the indicated trajectories to equip them with their respective rates. **c,** The additional second harmonic of the two-colour field modulates kinetic energies (colour) and rates (thickness) of the trajectories. This modulation depends markedly on the relative phase (compare the four panels). The overall yield is minimal (maximal) for a relative phase $\phi_{rel} = 0$ ($\phi_{rel} = \pi$) following the reduction (enhancement) of the combined field strength $|E_{\omega+2\omega}|$ around the instance of birth, as indicated by the field amplitudes beneath. Lowest (highest) final energies are realized for the phase $\phi_{rel} = \pi/2$ ($\phi_{rel} = 3\pi/2$), owing to the decreased (increased) difference in vector potentials $|\Delta A_{\omega+2\omega}|$ between birth and rescattering region (see areas of grey shaded regions and Methods for details).

For atoms and molecules in the gas phase, this has been done with utmost success, including the measurement of the electron emission durations as well as the timing of the recombination with attosecond precision[13,14]. A central trick employed in such measurements on atomic and molecular systems is the controlled distortion of the electron trajectory by an adjustable transverse electric field or field-induced molecular orientation, which allows controlling if a given electron trajectory re-encounters or misses the parent matter.

**Two-color modulation spectroscopy**

As the yield of high harmonic generation is small and trajectory selection via polarization fails due to the dominant normal component of optical nearfields at metallic nanostructures and solids, none of the above gas-phase techniques is applicable to solids, in particular not to needle tips. Here we report an alternative technique that allows capturing the photoemission dynamics using two-colour modulation spectroscopy (TCMS) with linearly polarized fields consisting of strong near-infrared pulses superimposed with their weak second harmonic.

The concept of TCMS is sketched in Fig. 1: Two-colour fields impinge on the needle tip as shown in Fig. 1a; the electron behaviour in the most intense half-cycle (highlighted) is of central importance. The strong fundamental field alone launches electron trajectories with and without rescattering, as evident from Fig. 1b.

Each trajectory is characterized by a rate (line width) resulting from the tunnelling probability and a terminal energy (colour) reflecting the field-driven electron motion. In the presence of the additional second harmonic field, both the emission probability[26] as well as the trajectories' terminal energy[27] are modified and vary in a characteristic way as function of the relative phase $\phi_{rel}$ of the two-colour field. This holds in particular for the backscattering electrons, which form the famous rescattering plateau[16]. Figure 1c illustrates the selective impacts on the trajectories of elastically backscattering electrons, which, for example, show the highest terminal energy or maximal overall yield for very different values of the relative phase (see figure caption).

Photoelectron spectra formed by all launched trajectories are recorded as function of the relative phase of the two-colour field and inherit the trajectories' energy and yield modifications as characteristic relative phase and energy-dependent spectral features[28]. Key to resolving the underlying sub-cycle emission dynamics is the precise matching of the measured modulation features with solutions of the time-dependent Schrödinger equation[27] (TDSE, cf. Supplementary Information), which are then projected onto classical trajectories. The photoemission rate (yield per birth time duration) determined from the projected quantum wavepacket dynamics (background grey-level plots in Figs. 1 b,c) therein acts as a temporal gate function for the trajectories and provides quantitative insights into both the quantum mechanical emission duration (timing of the tunnelling process) and the wavepacket dynamics after the electron is born into the nearfield. Central to the method is the simultaneous quantitative characterization of the nearfield amplitudes before the matching procedure by using the relative phase-dependent modulation of the cut-off energy in the rescattering plateau without any prior knowledge of the local optical field enhancement. We emphasize that the latter is otherwise difficult to measure[29].

**Spectral modulation signatures**

In the experiment, we employ few-cycle two-colour laser pulses to drive electron emission from a tungsten needle tip (cf. Fig. 1a). The linearly polarized two-colour field consists of 9 fs fundamental pulses centred at 1560 nm (~2 cycles) from an Erbium-doped fibre laser[30] and their second harmonic with 8 fs duration (~3 cycles) centred at 780 nm (Methods). The two-colour field is linearly polarized along the tip axis, tightly focused onto the tip apex (radius of curvature of 15 nm) and has nearfield peak field amplitudes of $E_\omega = 7.54\,\text{Vnm}^{-1}$ and $E_{2\omega} = 1.43\,\text{Vnm}^{-1}$ for the fundamental and the second harmonic deduced from independent single-color measurements (Methods) resulting in a nearfield admixture $\alpha = E_{2\omega}/E_\omega$ of 19%.

Figure 2a shows a measured relative phase-averaged electron energy spectrum (orange), which is matched well by the numerical solution (blue) of the TDSE. The spectra display the hallmarks of field-driven electron dynamics: a direct electron emission feature up to about $3.4\,\text{eV} \approx 2U_p$ and a nearly flat plateau associated with elastic backscattering with a cut-off at $17\,\text{eV} \approx 10U_p$ (orange circle, Methods) followed by an exponential roll off. Here, the ponderomotive energy $U_p$=1.71 eV is associated with the fundamental nearfield.

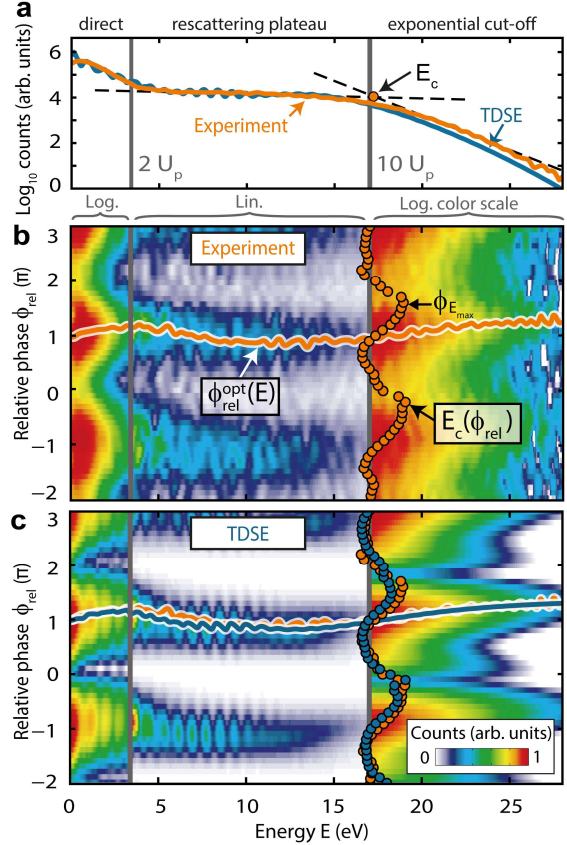

**Figure 2| Photoelectron energy spectra. a,** Measured relative phase-averaged photoelectron spectrum (orange graph) and as predicted by TDSE simulations (blue). The exponential decrease below 3.4 eV ~ $2U_p$ represents direct electron emission. It is followed by a backscattering plateau and an exponential cut-off at around 17 eV ~ $10U_p$ resulting from elastic rescattering. The cut-off energy $E_c$ (orange circle, Methods) is quantified via the intersection of linear fits in this plot (dashed lines) of plateau and cut-off region. **b,c,** False-colour maps of photoelectron spectra measured (b) and predicted by TDSE (c) as function of the relative phase $\phi_{rel}$ of the two-colour field. For clear visualization, we use logarithmic colour scales in the direct and cut-off domains and a linear scale in the plateau region, each normalized individually. The phase-dependent cut-off energies $E_c(\phi_{rel})$, obtained like in panel a, are indicated as coloured circles. Solid orange (experiment) and blue (simulation) graphs indicate the energy-dependent optimal phase $\phi_{rel}^{opt}(E)$ of maximum electron emission rate at a given electron energy $E$. We note that multiphoton peaks are less pronounced in the experiment due to limited spectrometer resolution. The orange graph and circles in panel c are copied from panel b to show the almost perfect match between simulation and experiment, notably in the cut-off part, which is essential to attain quantitative insights.

Figures 2b,c demonstrate the relative phase-dependent evolution of the measured and simulated spectra. They display strong variations of both the spectral emission rate and the cut-off energy with relative phase, representing our central observables for the characterization of both the electron emission dynamics as well as the nearfield amplitudes: From these spectra, we extract the optimal relative phase $\phi_{\text{rel}}^{\text{opt}}(E)$ defined by the maximal rate for a given electron energy $E$ (orange and blue curves in Fig. 2b,c; details in Extended Data Fig. 1) as well as the relative phase-dependent cut-off energy $E_c(\phi_{\text{rel}})$ (Methods). Most importantly, the simulation reproduces all main features of the experimental data very well, in particular the optimal relative phase and the cut-off energy, even though correlation effects as well as the electronic band structure of the tip material beyond the work function $W$ and the Fermi energy are not contained in the TDSE model. This important agreement justifies the use of the single active effective electron model, which represents a key result of this study.

**Trajectory analysis**
Although the TDSE results match the experimental features well, they do not offer direct insight into the physical processes and their timings. To extract the physics we link the quantum simulations with classical trajectories in the spirit of the seminal three-step model[11] generalized to two-colour fields[31]. For this, the gradient of the nearfield near the tip surface is neglected such that an effective local vector potential $A(t)$ depending on time $t$ can be used. In this simple-man's model (SMM) with classical trajectories, the final momenta of directly emitted and elastically backscattered electrons[16] respectively read $p_d = -eA(t_i)$ and $p_{\text{res}} = -e[2A(t_{\text{res}}) - A(t_i)]$ and depend on the electron's birth time $t_i$ and the rescattering time $t_{\text{res}}(t_i)$ with $e$ being the elementary charge. The vector potential of the two-colour field $A(t) = A_\omega(t) + A_{2\omega}(t, \alpha, \phi_{\text{rel}}) \equiv A_{\omega+2\omega}(t)$ and the rescattering time $t_{\text{res}}$ depend on the relative phase $\phi_{\text{rel}}$ and field admixture $\alpha$. Figure 3a displays representative examples of the resulting final kinetic energies for $\phi_{\text{rel}} = 0.5\pi$ and $\phi_{\text{rel}} = 1.5\pi$ versus birth time $t_i$. The impact of the second harmonic is most drastically visible in the modification of the peak rescattering energy (red and blue square). For sufficiently small field admixtures $\alpha$, the amplitude of the peak energy modulation with $\phi_{\text{rel}}$ can be shown to be linear in $\alpha$, with a classical energy maximum at $\phi_{E_{\max}}^{\text{SMM}} \approx 1.5\pi$ (Methods, corresponding trajectories in lower right panel of Fig. 1c).

TDSE simulations confirm the linear scaling of the modulation depth with $\alpha$ but yield a lower scaling factor than the classical prediction, which is mainly attributed to the averaging over the full set of (quantum) trajectories (Extended Data Fig. 2b). The phase of maximal cut-off energy, spectral shape and optimal phase are robust against changes of the admixture (Extended Data Fig. 2a,c,d). Using the proportionality factor from the TDSE and the clear link between admixture and cut-off modulation depth established by the SMM allows us to read off the field admixture present in the experiment. Together with the known fundamental field strength we recover $E_{2\omega} = 1.64$ V nm$^{-1}$ for the weak second harmonic field strength without prior knowledge, which is in good agreement with independent single-color measurements as provided earlier (Methods).

Extending the birth time-dependent final kinetic energies from Fig. 3a to all relative phases results in Fig. 3b. Here, we recover the modulation of the peak energy (maximum located again at red square) and find a slight but notable variation of the associated emission time (black curve). This final energy landscape for the precisely determined admixture completes the propagation aspect of our trajectory analysis, which we now combine with corresponding birth time-dependent rates to reveal the characteristics of the optimal phase $\phi_{\text{rel}}^{\text{opt}}(E)$ (Fig. 3c).

**Linking the optimal phase to rates**
The birth time and relative phase-dependent emission rate is obtained from TDSE simulations (Fig. 3d, parameters corresponding to Fig. 2c) by projecting out the contributions of the bound and transiently polarized electron population from the time-dependent wavefunctions (Methods). To explain the characteristic evolution of the optimal phase with energy we inspect three spectral regions within the direct (I), plateau (II) and cut-off (III) part of the electron spectrum marked in Fig. 3c. Areas in the birth-time versus relative-phase diagram in Fig. 3b that contribute to the energy domains I – III are marked as grey shaded areas and labelled accordingly. Crucially, only electrons from the areas I, II or III in Fig. 3b can contribute to the direct (I), plateau (II) or cut-off part (III) in Fig. 3c, respectively. Hence, these areas can be interpreted as spectrally selective masks in the photoemission rate map in Fig. 3d (shaded areas, repeated there).

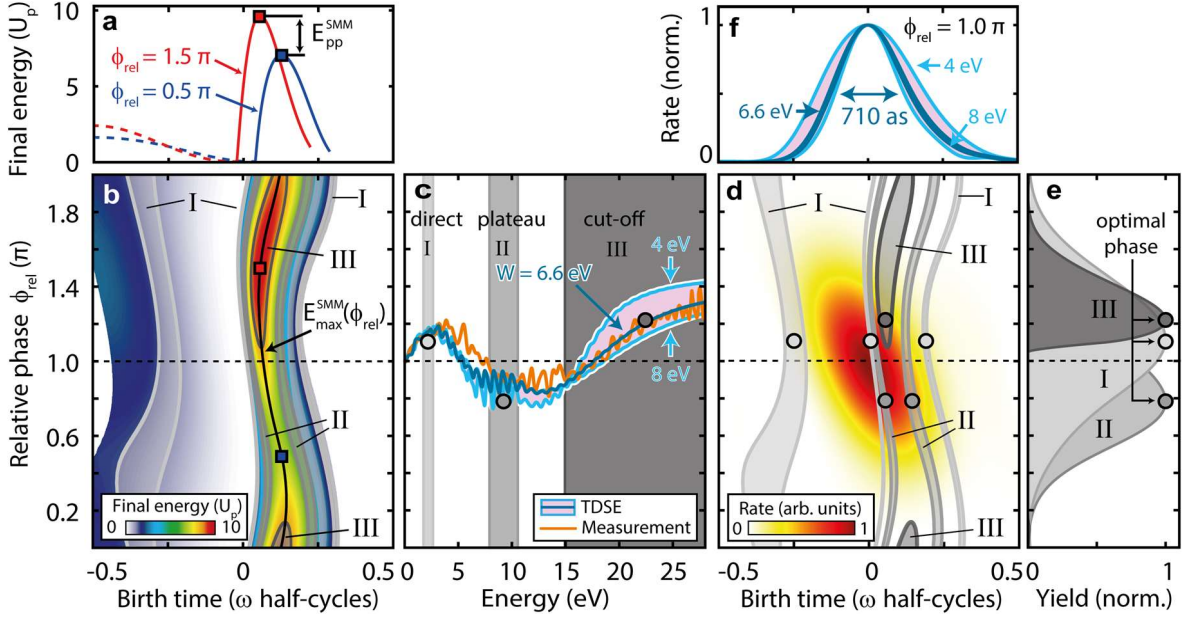

**Figure 3| Reconstructing trajectory and emission dynamics. a**, Final energies of classical trajectories for directly emitted (dashed) and backscattered (solid) electrons as function of birth time for two relative phases as indicated. Maximum energies are marked by red and blue squares. **b**, Final energies as function of relative phase and birth time (colour code) and maximum energy $E_{max}^{SMM}(\phi_{rel})$ (black contour connects maximum energies, squares like in panel a). **c**, Optimal phases extracted from the experiment (orange curve) and TDSE simulation (dark blue, work function $W = 6.6$ eV) from Fig. 2c. Light blue curves and purple shaded area visualize TDSE results with work functions ranging from 4 – 8 eV. Three characteristic spectral regions of direct emission (I), in the rescattering plateau (II) and cut-off domain (III) are marked in grey. Emission times and relative phases contributing to these domains are indicated by respectively coloured areas I, II, III in panel b. The optimal phase in the sensitive cut-off part almost perfectly matches to our TDSE simulations for $W = 6.6$ eV. We attribute the deviations in the direct and plateau domain to stronger interference structures and the idealized pulse shapes within the TDSE simulations. **d**, Instantaneous emission rate (colour code) extracted from the TDSE with $W = 6.6$ eV together with spectral regions I-III reinserted from panel b. **e**, Phase-dependent yields obtained by integrating the rates in panel d along the time axis belonging to the respective domains. Grey circles mark the relative phases of maximum yield and are inserted in panels c and d for comparison. **f**, Instantaneous rate for $\phi_{rel} = \pi$ (dashed line in panel d) for different work functions (blue curves, as indicated). The best matching work function $W = 6.6 \pm 0.3$ eV constrains the emission duration (FWHM) to $710 \pm 30$ as.

The resulting energy region-specific yield (rates integrated over birth time) as function of relative phase (Fig. 3e) shows maxima that coincide well with the optimal phase values in experiment and TDSE simulation, as indicated by the dots in Fig. 3c. The relative phase-dependent overlap of the region-specific masks in Fig. 3d and the photoemission rate hence determine the physics encoded in the optimal phase profile.

More precisely, for directly emitted electrons (I), the maximal yield emerges slightly above $\phi_{rel} = \pi$ (maximal emission rate), where high rates and large overlap of the spectral masks I and the photoemission distribution is found in Fig. 3d. The maximum yield in the plateau domain (II) is reached when both masks II are inside the high-rate region, as realized slightly below $\phi_{rel} = \pi$ (Fig. 3d). The cut-off domain (III) shows qualitatively different features due to the isolated island-like structure of the corresponding spectral mask III centred around $\phi_{rel} \sim 1.5\pi$ (Fig. 3b), with all significant contributions located well above the phase for maximal emission rate. Here, the competition between spectral masks and emission rate results in maximal yield between their peak values, namely at $\phi_{rel} = 1.22\pi$ (Fig. 3 d,e). This aspect highlights the notion that a photoemission distribution more localized in the relative-phase versus birth-time plot will shift the yield towards the peak emission rate. Hence, the isolated island's spectral mask structure of energy domain (III) explains the sensitivity of the optimal phase in the cut-off domain to the temporal width, and thus to the order of the non-linearity of the emission.

**Extracting the emission duration**

For larger values of the work function, that is, a higher non-linearity, the temporal localization of the emission rate becomes more pronounced, as expected (Fig. 3f). This fact leads to a systematic shift of the optimal phase towards $\phi_{rel} = 1.0\pi$ within both the TDSE (Fig. 3c) and our combined semi-classical trajectory analysis (Extended Data Fig. 3). Matching the predicted cut-off domain of the optimal phase profile to the experimental data thus allows us to specify the duration of the photoemission with high precision. The best match is found for a comparably large work function of $W = 6.6 \pm 0.3$ eV, reproducing previous observations of a significant work function increase in such experiments[9] and confirmed by complementary independent measurements (Supplementary Information). For the parameters of our proof-of-principle study (intensities, wavelengths, and work function), the resulting emission duration is $710 \pm 30$ as with an error bar inherited from the work function (detailed analysis and discussion in Methods, time-bandwidth considerations discussed in Supplementary Information) – here for the first time determined at the surface of a metallic solid. Although the exact value of the emission duration may change for other parameters, our analysis in the Supplementary Information suggests that our metrology remains robust.

**Conclusion**
We have presented Two-colour modulation spectroscopy (TCMS), allowing us to uncover the sub-cycle strong-field emission and propagation dynamics from a solid needle with attosecond precision previously reserved for atomic physics experiments. TCMS is applicable to any strong-field physics target, ranging from nanoplasmonic surfaces to well-studied atomic and molecular targets, and works with any wavelength provided backscattering is present. We note that such quantitative insights could so far not be achieved by modulating the carrier-envelope phase of few-cycle waveforms, although our here-developed metrology should also be applicable in this case (see Supplementary Information). Our work further paves the way for petahertz (or lightwave) electronics by specifying a quantum-limited bandwidth equalling $1.41 \pm 0.06$ PHz derived from the inverse of the emission duration for our parameters. Within nanoplasmonics TCMS will help to map smallest nearfields, investigate the role of excited states in the ultrafast emission process and provide tailored broadband and ultrashort electron pulses as a probe. Finally, it will be essential to quantify the emission duration at metal surfaces to properly assess fundamental limits caused by Pauli-blocking[18,19] and Coulomb interactions[18,20,21,32] when multiple electrons are emitted within an ultrashort time window at nanometre distances, with direct relevance to time-resolved electron microscopy.

*Acknowledgements*

This work has been supported in part by the European Research Council (Consolidator Grant NearFieldAtto, Advanced Grant AccelOnChip) and the Deutsche Forschungsgemeinschaft (priority program SPP 1840 QUTIF). T.F. acknowledges financial support from the Deutsche Forschungsgemeinschaft within the Heisenberg programme (IDs: 315210756, 398382624, 436382461) and via CRC 1477 "Light-Matter Interactions at Interfaces" (ID 441234705). A.Li. and A.Le. acknowledge funding by DFG, ID 425217212 – SFB 1432. We thank N. Dudovich, M. Krüger and Chr. Lemell for discussions.

## Methods

**Experimental setup**

In our setup (Extended Data Fig. 4a), pulses from an Erbium-doped fibre laser with an initial pulse duration of 74 fs at 100 MHz repetition rate and central wavelength of 1560 nm are shortened to few-cycle duration. To achieve this, the pulses are pre-chirped by a silicon prism compressor and launched into an optical fibre composed of a single-mode section and a millimetre short, but highly non-linear Germanium-doped section, where spectral broadening and self-compression take place[30]. The few-cycle fibre output is collimated by a 90° off-axis parabola (f = 5 mm) and expanded by a 1:3 reflective telescope (not shown in sketch), before it is focused into a 0.1 mm thick BIBO crystal to generate its second harmonic. The 90° off-axis parabolas with 15 mm focal length are used for the focusing and re-collimation in the second harmonic generation stage. A dichroic beam splitter transmits the fundamental (1000 nm – 2000 nm) and reflects the second harmonic (600 nm – 1000 nm) spectral components at the input of a dichroic Mach-Zehnder-type interferometer. The two spectral domains are individually dispersion compensated by chirped mirrors in each arm and can be attenuated with neutral density filters. Any residual leakage of the fundamental spectrum in the second harmonic interferometer arm and vice versa is removed by polarization filters. A variable delay is added to the fundamental pulses and a 1:1 telescope matches the beam divergence before both beams are combined at a dichroic beam splitter defining the exit of the interferometer. The resulting collinear two-colour field is tightly focused onto the apex of a tungsten needle tip within a UHV-vessel (pressure $\sim 1 \times 10^{-10}$ hPa) with beam waist radii ($1/e^2$ intensity radius) $w_\omega = 2.1 \pm 0.2$ μm for the fundamental and $w_{2\omega} = 2.0 \pm 0.2$ μm for the second harmonic field. The pulse durations at the interaction point are $\tau_\omega = 9 \pm 1$ fs for the fundamental and $\tau_{2\omega} = 8 \pm 1$ fs for the second harmonic field, both defined by the full width at half maximum of the intensity envelope obtained via frequency resolved optical gating (FROG[33]) taking the propagation distance and vacuum window into account. The pulse energy from shot-to-shot and over long time scales fluctuates with a standard deviation of less than 0.6% from its means value for the fundamental and second harmonic. A continuous wave laser ($\lambda = 532$ nm) is coupled in by a thin fused silica plate and propagates collinearly with both colours. Its transmitted part at the second dichroic beam splitter is used to actively stabilize and shift the relative phase between the two colours using a Pancharatnam phase-lock[34]. The integrated relative phase noise is 70 mrad (standard deviation) determined by major noise components from 1 kHz to 0.1 Hz, which we attribute to mechanical vibrations and airflow.

The nanometric needle tip is wet-etched from a monocrystalline tungsten (310) wire. The ring counting method[35] (Extended Data Fig. 4b) during *in-situ* field-ion microscopy (FIM) allows us to infer an apex radius of 15 nm before the tip was mounted on a 3D translation stage in the experimental chamber. The same FIM image (Extended Data Fig. 4c) is reproduced in the experimental chamber using a microchannel plate (MCP) in front of the tip. Field emission occurs for a negative bias voltage of $-390$ V applied to the tip matching the location of the (310) plane. Laser emitted electrons at a reduced negative bias of $-100$ V in Extended Data Fig. 4e show the same spatial distribution as field emitted electrons in Extended Data Fig. 4d.

To record photoelectron spectra, the MCP is replaced by a retarding field spectrometer also used in Ref. [9]. The tip is grounded, and the spectrometer entrance is biased at $+50$ V resulting in a static field of $\sim 0.4$ GVm$^{-1}$ at the tip apex, which is more than one order of magnitude smaller than the applied optical field strengths. The post-processing of the spectra is described in the supplementary information of Ref.[9]. The experimental data was taken in relative phase steps $0.1\pi$ of using the Pancharatnam phase-lock. The count rate stability is similar for all relative phases with a standard deviation of $\sigma_{N=1} = 4.0$ % for 10 ms time bins. Each final spectrum is the average of 8 individual, consecutive spectra reducing the standard deviation to $\sigma_{N=8} = 1.4$ %. Each final spectrum requires a measurement duration of 5 minutes. The relative phase in the experiment can only be determined up to a global offset, which we chose to match the TDSE simulations. For the matching, we use the optimal phase in the energy range from 0 to 3 eV in the phase-resolved photoelectron spectra. In this spectral range, the simulated optimal phases as function of the work function deviate least among each other and match the experiment closest in shape. This procedure turned out to be more precise than only matching the maximum positions in the total yield.

**Determination of near-field strengths, field enhancement factors and field admixture**

To achieve the goal of determining the enhanced near-field strengths and the respective field enhancements for the individual components of the two-colour field, we employ single colour measurements and simulations to extract the respective cut-off energies of recollision electrons. We define these cut-off energies by the intersection of linear fit functions within electron spectra in a semi-logarithmic representation as indicated in Fig. 2a and Extended Data Fig. 5a. The first function is fitted within the plateau domain and the second function within the cut-off domain. This definition makes the cut-off position independent of the overall count rate, which corresponds to a vertical shift of the spectra in the semi-logarithmic representation.

For the fundamental, we relate our cut-off definition to the ponderomotive energy $U_\text{p} = \frac{e^2 E_\omega^2}{4m\omega^2}$ defined for the enhanced near-field strength $E_\omega$ by using electron spectra from TDSE simulations as shown in Extended Data Fig. 5a. Here, $e$ is the elementary charge, $m$ the electron mass and $\omega$ the angular frequency of the fundamental field. To automate the cut-off determination, we define the plateau domain as the energy interval from $4\,U_\text{p}$ to $8\,U_\text{p}$ and the cut-off domain as the interval from $11\,U_\text{p}$ to $14\,U_\text{p}$, where the ponderomotive energy is known in the simulations. As shown in Extended Data Fig. 5b, the resulting cut-off energies scale linearly as function of the ponderomotive energy (see blue circles) but slightly deviate from the classical[36] $10\,U_\text{p}$ and quantum orbit[37] $10\,U_\text{p} + 0.5W$ rules with the work function $W$ (corresponding to the binding energy $I_\text{p}$ in Ref.[37]). We take these systematic deviations into account by directly relating the cut-offs to $U_\text{p}$ using a linear fit function (solid blue line).

For the experimental spectra we proceed analogously but, as the local field enhancement and the respective ponderomotive energy are initially unknown, employ an iterative scheme where we manually define initial intervals in the plateau and cutoff regions. The ponderomotive energy is determined from the cut-off energy extracted via fitting the spectra in these intervals and compared with the theoretical estimation. Then, the fully automatic evaluation is performed to refine the cut-off energy and provides the final value of $U_\text{p}$. To evaluate the cut-offs in the phase-resolved spectra shown in Fig. 2b,c, we use a single colour ponderomotive energy of $U_\text{p} = 1.71$ eV (as determined in the following) to define the initial fitting intervals.

For the fundamental field we obtain $U_\text{p} = 1.71 \pm 0.03$ eV, which directly leads to a near-field strength of $E_\omega = 7.54 \pm 0.07\,\text{Vnm}^{-1}$ (orange circle in Extended Data Fig. 5b). We obtain the field enhancement by comparison with the strength of the incident field. For an average power of $P_\omega = 50.4 \pm 2.5$ mW (as used for the fundamental in the two-colour experiment), we obtain an incident field strength of $E_\omega^\text{inc} = 1.51 \pm 0.19\,\text{Vnm}^{-1}$, where the absolute error

$$\Delta E_\omega^\text{inc} = \sqrt{\frac{1}{4}\left(\frac{\Delta P_\omega}{P_\omega}\right)^2 + \frac{1}{4}\left(\frac{\Delta \tau_\omega}{\tau_\omega}\right)^2 + \left(\frac{\Delta w_\omega}{w_\omega}\right)^2}\, E_\omega^\text{inc}$$

includes relative errors in the pulse duration $\Delta\tau_\omega/\tau_\omega = 10\%$, the beam waist $\Delta w_\omega/w_\omega = 10\%$ and the incident power $\Delta P_\omega/P_\omega = 5\%$. Relating the near-field strength to the incident field strength yields a field enhancement factor of $\gamma_\omega^0 = 5.0 \pm 0.6$.

To further validate our method, we sweep the incident power in the experiment and determine cut-offs from the spectra with the previously described iterative routine. The obtained cut-offs as function of the local ponderomotive energy $U_\text{p}$ (all scaled with respect to $U_\text{p} = 1.71$ eV at $P_\omega = 50.4$ mW including the uncertainty in the power measurement) align well with the TDSE prediction (compare black and blue circles in Extended Data Fig. 5b).

For determining near-field strength and field enhancement factor for the weaker second harmonic we proceed analogously. However, as the maximum achievable second harmonic field is too weak to directly drive electron rescattering within our setup, we measure rescattering spectra with a Titanium:Sapphire oscillator with nearly identical wavelength and pulse duration instead to determine the field enhancement factor of $\gamma_{2\omega}^0 = 3.5 \pm 0.35$.

For an incident power of $P_{2\omega} = 2.8 \pm 0.3$ mW, corresponding to an incident field strength of $E_{2\omega}^\text{inc} = 0.41 \pm 0.05\,\text{Vnm}^{-1}$ of the frequency-doubled pulses from the Erbium laser system and considering the field enhancement factor derived above, we obtain a near-field strength of $E_{2\omega} = 1.43 \pm 0.23\,\text{Vnm}^{-1}$. The relative errors in the field enhancement factor and the beam waist of the second harmonic mainly determine the absolute error

$$\Delta E_{2\omega} = \sqrt{\frac{1}{4}\left(\frac{\Delta P_{2\omega}}{P_{2\omega}}\right)^2 + \frac{1}{4}\left(\frac{\Delta \tau_{2\omega}}{\tau_{2\omega}}\right)^2 + \left(\frac{\Delta w_{2\omega}}{w_{2\omega}}\right)^2 + \left(\frac{\Delta \gamma_{2\omega}}{\gamma_{2\omega}}\right)^2} E_{2\omega}$$

For the two input powers of the individual colours stated above this leads to a near-field admixture $\alpha = \frac{E_{2\omega}}{E_\omega} = (19 \pm 3)\,\%$.

**Trajectory analysis**
In this section, we derive analytical expressions for the final kinetic energies of directly emitted and recollision electrons in a two-colour near-field of a nanometric tip following the famous Simple Man's Model[11] (SMM) of strong-field physics. Thereto, electrons are described with classical trajectories, launched at rest at the surface ($x_0 = 0$, $v_0 = 0$), propagated under the impact of the two-colour field[31,38] and considered to scatter elastically when returning to the surface (velocity $v \to -v$).

While tracking electron trajectories in a spatially inhomogeneous near-field requires numerical integration of the classical equations of motion, considering a homogeneous field profile $E(t)$ depending on time $t$ and using $E(t) = -\mathrm{d}/\mathrm{d}t\, A(t)$ allows an explicit treatment in terms of the vector potentials of the two individual field components

$$A_\omega(t) = -\frac{E_\omega}{\omega}\int_{-\infty}^{\omega t} f_\omega(\phi)\cos(\phi + \phi_{ce})\,\mathrm{d}\phi \quad (1)$$

$$A_{2\omega}(t) = -\frac{\alpha E_\omega}{2\omega}\int_{-\infty}^{2\omega t} f_{2\omega}(\phi+\phi_{rel})\cos(\phi + 2\phi_{ce}+\phi_{rel})\,\mathrm{d}\phi. \quad (2)$$

with the carrier-envelope phase $\phi_{ce}$. The field envelopes as function of phase $\phi$ are defined as

$$f_{\omega/2\omega}(\phi) = \exp\left(-2\ln 2\left(\frac{\phi}{\tau_{\omega/2\omega}^{cycle}}\right)^2\right), \quad (3)$$

where the pulse durations $\tau_{\omega/2\omega}$ are related to their durations in cycles $\tau_\omega^{cycle} = \omega\tau_\omega$ and $\tau_{2\omega}^{cycle} = (2\omega)\tau_{2\omega}$. We proceed with the analysis by considering the normalized vector potentials $\tilde{A}_\omega(t) = \frac{\omega}{E_\omega}A_\omega(t)$ and $\tilde{A}_{2\omega}(t) = \frac{2\omega}{\alpha E_\omega}A_{2\omega}(t)$ to express the combined two-colour vector potential $A(t) = \frac{E_\omega}{\omega}\left[\tilde{A}_\omega(t) + \frac{\alpha}{2}\tilde{A}_{2\omega}(t)\right]$.

The final momenta of direct electrons launched at time $t_i$ are $p_d = -eA(t_i) = -\frac{eE_\omega}{\omega}\left[\tilde{A}_\omega(t_i) + \frac{\alpha}{2}\tilde{A}_{2\omega}(t_i)\right]$, following from integrating Newton's equation of motion. This translates to the final kinetic energy of directly emitted electrons in units of the ponderomotive energy

$$E_d = \frac{p_d^2}{2m}$$
$$= 2U_p \times \left[\tilde{A}_\omega(t_i) + \frac{\alpha}{2}\tilde{A}_{2\omega}(t_i)\right]^2 \quad (4)$$

For electrons rescattering at time $t_{res}$ the final momenta read

$$p_{res} = -e(2A(t_{res}) - A(t_i))$$
$$= -\frac{eE_\omega}{\omega}\Big(2\tilde{A}_\omega(t_{res}) - \tilde{A}_\omega(t_i)$$
$$\quad + \frac{\alpha}{2}[2\tilde{A}_{2\omega}(t_{res}) - \tilde{A}_{2\omega}(t_i)]\Big)$$
$$= -\frac{eE_\omega}{\omega}\left(\Delta\tilde{A}_\omega + \frac{\alpha}{2}\Delta\tilde{A}_{2\omega}\right) \quad (5)$$

which allows us to express the final kinetic energies of rescattered electrons as

$$E_{res} = 2U_p \times \Big[\left(\Delta\tilde{A}_\omega\right)^2 + \alpha\left(\Delta\tilde{A}_\omega\,\Delta\tilde{A}_{2\omega}\right) + \frac{\alpha^2}{4}\left(\Delta\tilde{A}_{2\omega}\right)^2\Big] \quad (6)$$

Here, $\Delta\tilde{A}_\omega$ and $\Delta\tilde{A}_{2\omega}$ correspond to the momentum gains imposed by the individual field components and accumulated from the moment of birth $t_i$ to recollision at time $t_{res}$ and the following escape from the surface. We note that the vector potentials depend on the carrier envelope phase and the relative phase, which we omitted in our notation for the sake of readability.

Extended Data Figure 6a shows the final electron energy (black curves) as function of the birth time for a two-colour field with field admixture $\alpha = 20\%$. In contrast to atoms, electrons can only be emitted into vacuum for half-cycles where electric field vectors point into the metal. The highest energy for direct electron emission (dashed black curves) occurs during the central cycle (red shaded area) where the strongest vector potential is realized (cf. equation (4)). The fastest rescattering electrons originate from the cycle prior to the central cycle, as their rescattering time also coincides closely with the strongest vector potential, which dominates the final energy of rescattered electrons (cf. equations (5) and (6)). Varying the relative phase of the two-colour field enables us to calculate birth time and phase-dependent final kinetic energy maps as presented in Fig. 3b in the main text. Those allow us to inspect the relative phase-dependent modulation of the cut-off energy and extract the modulation depth as well as the phase maximizing the energy.

The modulation depth depends linearly on the field admixture (see solid blue curve in Extended Data Fig. 6b) and the maximal rescattering energy (Extended Data Fig. 6c) is found for a relative phase $\phi_{E_{max}}^{SMM} \approx 1.5\,\pi$ which is in good agreement with the experimental and TDSE results. This linear dependence and the relative phase for maximal energy gain are preserved even if the impact of the second harmonic on the rescattering time is neglected (dashed blue curve representing $t_{res}(t_i, \phi_{ce})$ in Extended Data Fig. 6b,c), which becomes evident in equation (6) when further neglecting the small $\alpha^2/4$ term. The energy modulation is therefore dominated by the modified momentum gain related to the relative phase-dependent vector potential of the second harmonic within the $\Delta \tilde{A}_\omega \Delta \tilde{A}_{2\omega}$ term in equation (6). We note that the impact of the carrier-envelope phase on the cut-off modulation depth and relative phase maximizing the rescattering energy is very small and thus neglected (see shaded areas in Extended Data Fig. 6b,c).

**Extraction of instantaneous ionization rates from TDSE simulations[39]**
To weight the classical trajectories, we derive time-dependent instantaneous ionization rates $R(t) = d/dt\,Y(t)$ related to the norm $Y(t) = \langle \Psi_{lib.}(x,t) | \Psi_{lib.}(x,t) \rangle$ for the liberated part $\Psi_{lib.}(x,t)$ of the full wavefunction $\Psi(x,t)$ (see Extended Data Fig. 7a and Supporting Information for propagation) located outside the metal, that is, for positions $x > 0$. A straightforward definition of the liberated part could be to consider only spectral components in the continuum obtained by projecting out the ground state $\Psi_{ground}(x)$ via

$$\Psi_{lib.}(x,t) = \Psi(x,t) - \langle \Psi_{ground}(x) | \Psi(x,t) \rangle \Psi_{ground}(x). \quad (7)$$

However, as clearly visible in Extended Data Fig. 7b, the resulting wavefunction still contains a transient polarization attributed to the excitation of resonances in the continuum $\Psi_{res}^{(i)}(x)$ (Extended Data Fig. 7c).

As these resonances lead to pronounced oscillations of the norm, but do not contribute to the emitted yield, we additionally project out the first $N$ resonances. This procedure aims at removing the reversible transient polarization components of the response. We choose $N$ such that oscillations in the yield due to reversible polarization are suppressed and such that the yield increases monotonously with time. The liberated part of the wavefunction is then defined as

$$\Psi_{lib.}(x,t) = \Psi(x,t) - \sum_{i=0}^{N} \langle \Psi_{res}^{(i)}(x) | \Psi(x,t) \rangle \Psi_{res}^{(i)}(x), \quad (8)$$

where we set $i = 0$ (i.e. $\Psi_{res}^{(0)}(x) = \Psi_{ground}(x)$) for the ground state and $i > 0$ for the continuum resonances. The continuum resonances $\Psi_{res}^{(i)}(x)$ originate from a series expansion of the bound state

$$\Psi_{pert}^{E}(x) = \Psi_{res}^{(0)}(x) + \lambda \Psi_{res}^{(1)}(x) + \lambda^2 \Psi_{res}^{(2)}(x) + \cdots \quad (9)$$

perturbed by a homogeneous and static electric field of strength $E$ for $x > 0$ as indicated in Extended Data Fig. 7c. For the unperturbed case we identify $\Psi_{pert}^{E=0}(x) = \Psi_{ground}(x)$. We now choose a set of $N$ increasing field strengths $E_1 < E_i < \cdots < E_N$ and compute each $\Psi_{pert}^{E_i}$ via imaginary time propagation. We note that we set the potential $xE_i = 0$ if $xE_i < -|W|$ during imaginary time propagation such that $\Psi_{pert}^{E_i}$ remains located within the binding potential $V_0$. The $i$-th resonance

$$\Psi_{res}^{(i)}(x) = \Psi_{pert}^{E_i}(x) - \sum_{j=0}^{i-1} \langle \Psi_{res}^{(j)}(x) | \Psi_{pert}^{E_i}(x) \rangle \Psi_{res}^{(j)}(x) \quad (10)$$

is found by projecting out all resonances for weaker perturbations $E_j < E_i$ from the current $\Psi_{pert}^{E_i}(x)$. After normalization the resonances form an orthonormal basis set for any static perturbation and approximate the transient polarization in the dynamical case.

The successive removal of resonances according to equation (8) is shown in Extended Data Fig. 8a-d and reduces the reversible transient polarization substantially. Extended Data Figure 8e shows the yield $Y$ (orange) and respective rate $R$ (blue) evaluated after removing all resonances up to $\Psi_{res}^{(3)}$. Note, that removing less resonances leaves oscillations in the yield and removing even more resonances hardly changes the resulting yield and rate. We define the emission duration as the FWHM of the rate $R$ as indicated.

Contributions of the emitted wavefunction that are driven back to the surface may interfere with the emission in consecutive cycles (see Extended Data Fig. 9a), leading to artificial oscillations of the norm and resulting in unphysical negative emission rates (see Extended Data Fig. 9c). To suppress this part of the wavefunction we truncate the laser field for distances $x > 10\text{Å}$ in the rate analysis only, which results in predominantly outgoing contributions as shown in Extended Data Fig. 9b. This allows us to extract a meaningful time dependent emission rate for all laser cycles, which is robust against small variations of the truncation distance. We used the truncated field for the rate analysis presented in the main text and Extended Data Figs. 7 and 8.

**Estimation of work function and relation to emission duration**

Our analysis and the simulation results in Fig. 3 show that, and explain why, the high-energy part of the optimal phase is sensitive to the emission time window. The latter itself is closely related to the work function at a given intensity. Determining the emission time window from the optimal phase requires two steps: Firstly, we determine the work function for which the best match between simulated and measured optimal phase is achieved. Secondly, we map the work function and its uncertainty onto corresponding emission durations.

By performing TDSE simulations with 0.1 eV step width in the work function we found a fully monotonous and smooth dependence for the optimal phase in the cut-off domain. For work functions exceeding 6 eV also all other domains hardly change. Thus, the optimal phase provides a one-to-one mapping onto a specific work function and intermediate values can be approximated by linear interpolation, which allows us to perform a continuous least-square optimization.

The best matching result for an estimated work function of $6.61 \pm 0.32$ eV is shown in Extended Data Fig. 10a, where the two sources of uncertainty are the errors in the optimal phase evaluation (on average $0.04\pi$, cf. Extended Data Fig. 1) and the offset error $\sigma_{\text{offset}} = 0.017\pi$ in the relative phase axis alignment of theoretical and experimental curves. The deviations in the plateau domain are systematic and not affected by the field admixture nor the work function. These deviations will be subject to further investigations, which are beyond the scope of this manuscript.

In order to map the work function to the width of the emission time window we inspect the instantaneous rate distributions (cf. Fig. 3d and Extended Data Fig. 3b,e) at $\phi_{\text{rel}} = \pi$ as function of the work function and calculate the emission duration (FWHM), see blue curve in Extended Data Fig. 10b. For the previously determined work function interval $6.6 \pm 0.3$ eV (grey area) the emission duration is $710 \pm 30$ as. Note that this analysis was performed for at $\phi_{\text{ce}} = \pi$, but we checked that the result is robust with respect to CEP-averaging (blue circles). Also, any kind of emission delay is included in the TDSE simulations used to match the experimental data. The interpretation in Fig. 3d does not change for small delays/shifts of the overall rate distribution.

*Method references*

*Extended Data Figures*

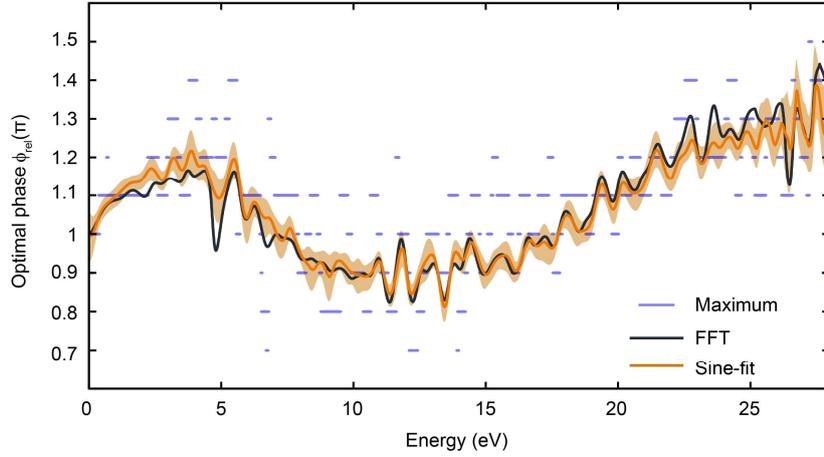

**Extended Data Figure 1| Different evaluation methods for the optimal phase**. For each energy $E$ we fit the rate with a cosine function and define the optimal phase as the phase maximizing the rate (shown here for the measurement). The semi-transparent orange band indicates the $1\sigma$ confidence interval of the fits with average error of $\pm 0.04\pi$. The considered relative phase interval extends from $-2\pi$ to $3\pi$. Alternatively, the optimal phase can be extracted by tracking the FFT-phase (black curve) or the maximum count rate (blue lines). In the latter method the resolution is restricted to discrete relative phase steps of $0.1\pi$.

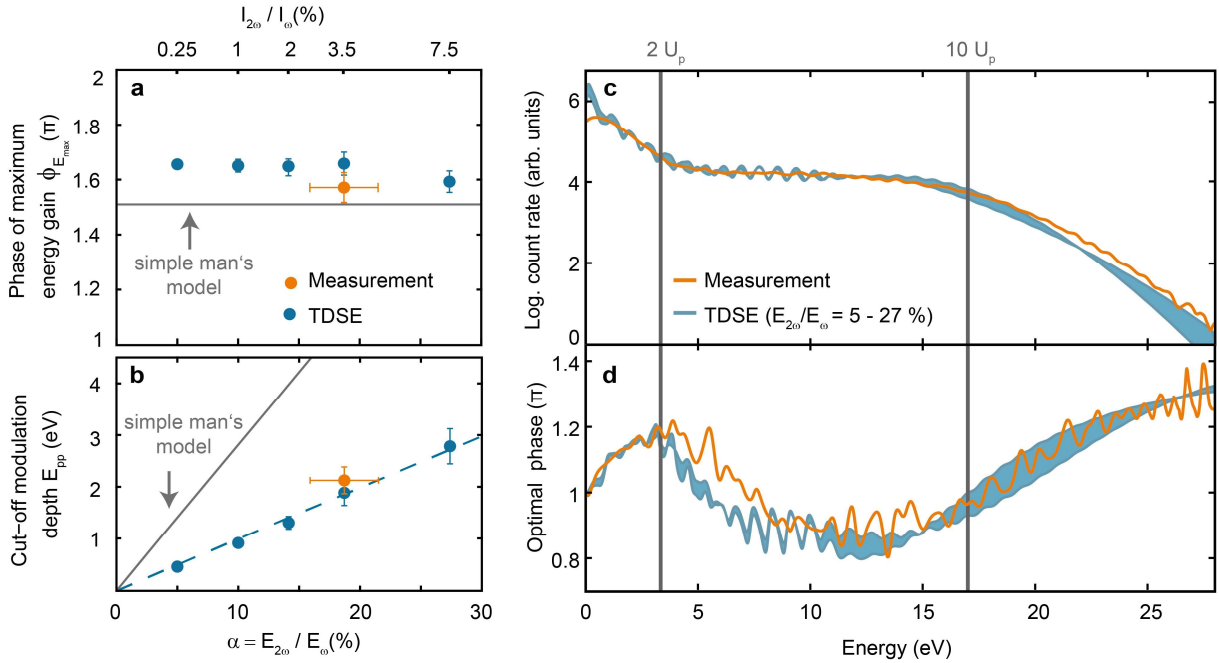

**Extended Data Figure 2| Impact of the field admixture. a,** Relative phases $\phi_{E_{max}}$ maximizing the cut-off energy and **b,** corresponding peak-to-peak cut-off modulation depths $E_{pp}$ as function of field admixture $\alpha$ (lower horizontal axis) or the respective intensity admixture (upper axis) extracted from TDSE simulations (blue symbols) and the experiment (orange symbols). Solid lines indicate the predictions of the simple man's model (SMM). The dashed line is a linear fit of the TDSE results. **c,** Phase-averaged photoelectron spectra of experiment (orange) and TDSE as blue shaded band containing all field admixtures from 5 to 27 %. **d,** Optimal phase for the same field admixture range and with the same colour code as in panel a. Clearly, the overall shape of the spectra (c) and optimal phase (d) hardly vary when the relative field strength of the second harmonic is varied in the large range of 5 to 27% of the fundamental field. Most importantly, this highlights the robustness of TCMS for extracting the attosecond emission time window via the optimal phase independent of the field admixture.

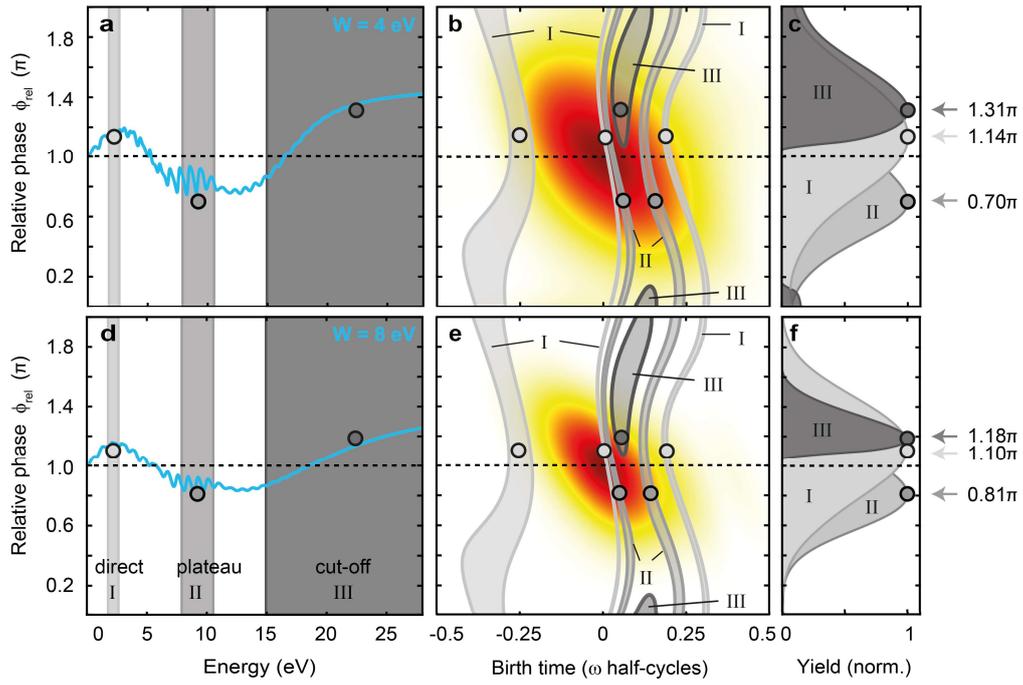

**Extended Data Figure 3| Relation between optimal phase and work function.** Optimal phases (a,d), birth time and relative phase-dependent instantaneous emission rates (b,e) and phase-dependent yields (c,f) as in Fig. 3 c-e, extracted from the TDSE simulation for work functions $W = 4$ eV (a-c) and $W = 8$ eV (d-f). Small/large work functions result in temporal broadening/confinement of the emission window (cf. Fig. 3f), which (following the argumentation in Fig. 3) shifts the optimal phase in the cut-off domain away from/towards $\phi_{rel} = \pi$. This is most clearly visible in the right-most column, where the spread of the optical phases for the three spectral regions is much larger for $W = 4$ eV (top row) than for $W = 8$ eV (bottom row). This fully explains the behaviour of the optimal phase and, reversely, allows us to extract the electron emission duration with high precision.

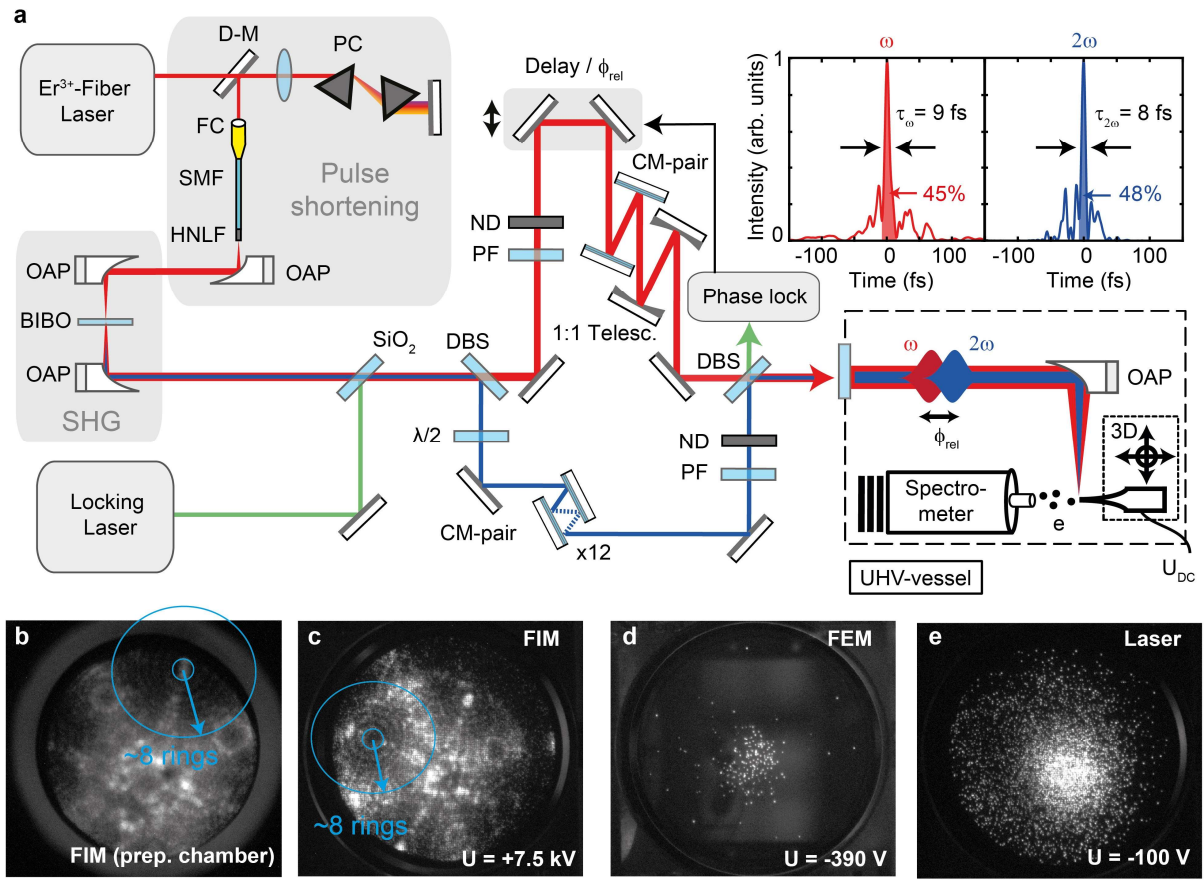

**Extended Data Figure 4| Experimental setup and tip characterization. a,** Experimental setup consisting of a pulse shortening stage, second harmonic generation (SHG) stage, dichroic Mach-Zehnder interferometer and UHV-vessel containing the needle tip and the electron spectrometer. Abbreviations: half-mirror (D-M), prism compressor (PC), fibre collimator (FC), single-mode fibre (SMF), highly non-linear fibre (HNLF), 90° off-axis parabola (OAP), bismuth borate crystal (BIBO), fused silica plate ($SiO_2$), dichroic beam splitter (DBS), half-wave plate ($\lambda/2$), polarization filter (PF), neutral density filter (ND) and chirped mirror (CM). Top right insets: pulse shapes, pulse durations and percentages of energy contained in the main peaks determined from frequency resolved optical gating (FROG) measurements of the fundamental (red) and second harmonic pulses (blue). **b,** Field ion microscopy image of the tip in preparation chamber. The ring counting method shows ~8 rings from the (110) to (211) pole corresponding to an apex radius of ~15 nm. **c,** Field ion microscopy image in experimental chamber. **d,** Field emission image for bias voltage of U = −390 V and **e,** laser emitted electrons for bias voltage of U = −100 V.

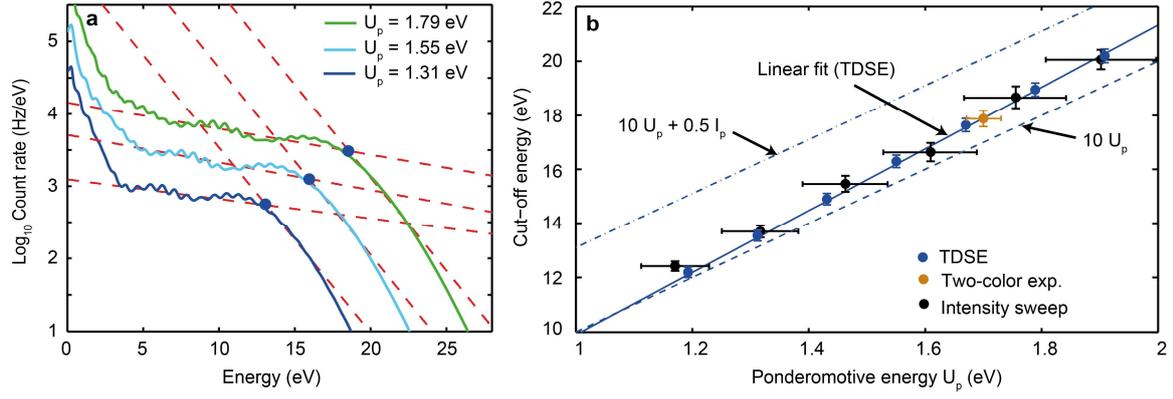

**Extended Data Figure 5| Cut-off extraction and relation to the ponderomotive energy. a,** Single-colour TDSE spectra for the fundamental field with increasing ponderomotive energy $U_p$. **b,** Cut-off energies evaluated from single-colour TDSE simulations (blue circles) and experimental electron spectra (orange and black circles) as function of the ponderomotive energy $U_p$. The cut-off position for the fundamental field strength used in the two-colour experiment is highlighted in orange. Both the cut-off positions from the simulated and the experimental data points depend linearly on the ponderomotive energy, but slightly deviate from the $10U_p$ and $10U_p + 0.5W$ rules (dashed lines).

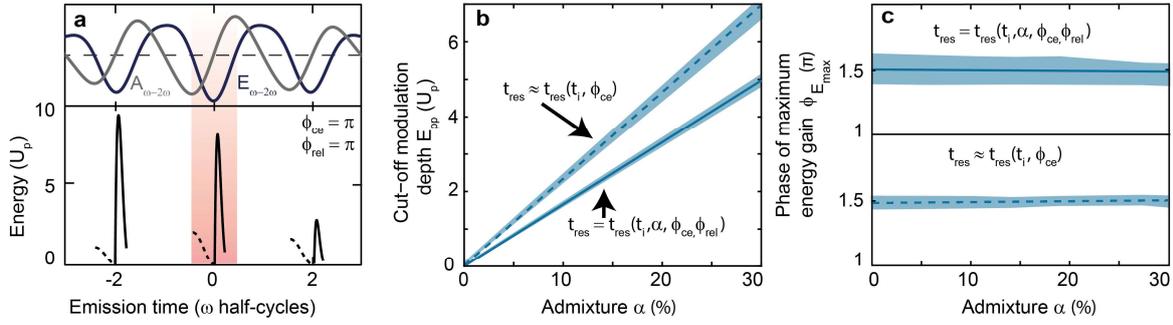

**Extended Data Figure 6| Trajectory Analysis,** Final kinetic energies (black curves) in units of the ponderomotive potential within the central cycles of a two-colour field (field and vector potential see top panel) obtained from SMM calculations at a field admixture of 20% (phases as indicated). **b,c,** Solid curves show the CEP-averaged peak-to-peak cut-off modulation depths $E_{pp}$ (b) and relative phases $\phi_{E_{max}}$ (c) resulting in maximal energies as function of field admixture as predicted by SMM. Dashed curves indicate respective results when neglecting the modification of the rescattering times due to the presence of the second harmonic (i.e. $\alpha$ and $\phi_{rel}$). Shaded areas indicate the small variations of the respective properties with the CEP.

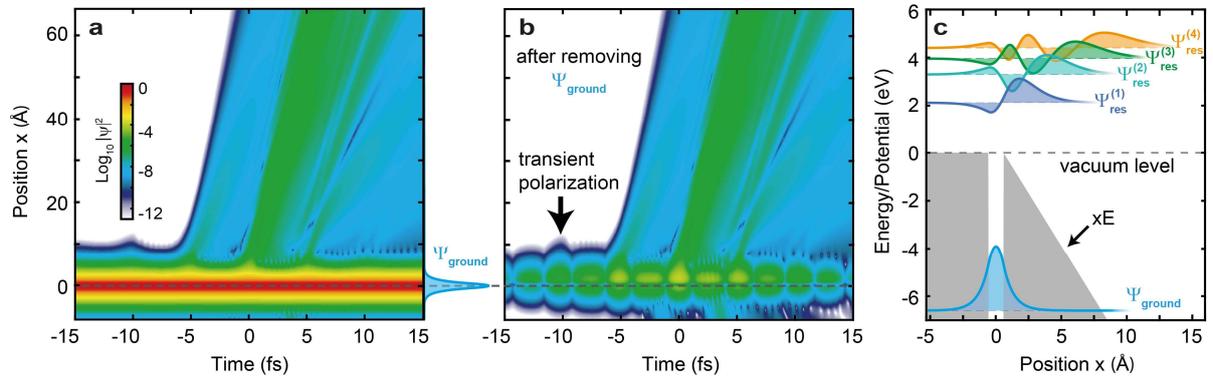

**Extended Data Figure 7| Time-dependent wavefunction and continuum resonances. a,** Probability density of the propagated wavefunction $\Psi$ in logarithmic colour-scale with ground state indicated on the right (blue). **b,** Probability density after projecting out the ground state. **c,** Binding potential perturbed by electric field $E$ (gray). The unperturbed ground state $\Psi_{ground}$ and first four continuum resonances $\Psi_{res}^{(1)}$ to $\Psi_{res}^{(4)}$ associated with the perturbed potential are indicated.

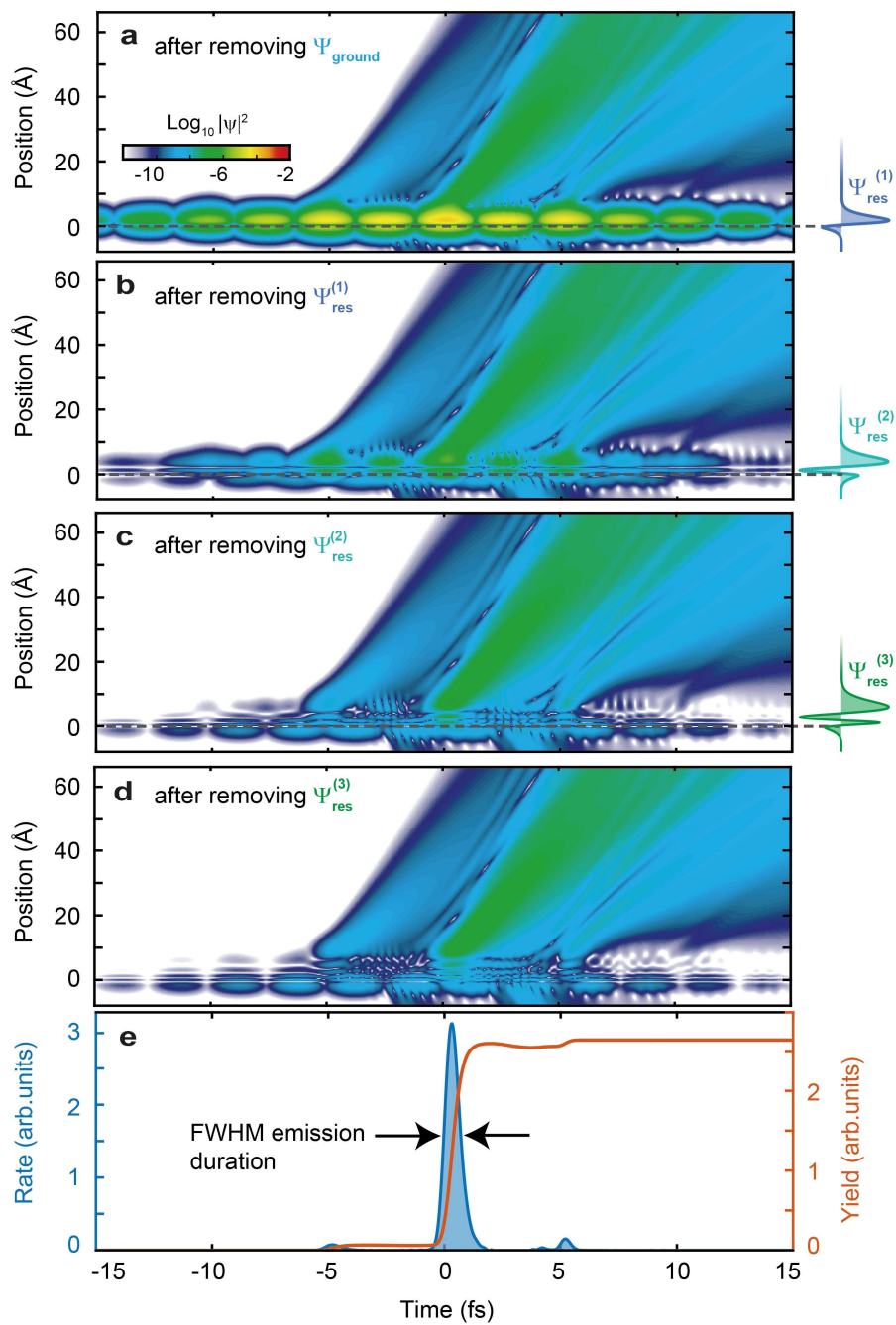

**Extended Data Figure 8| Successive removal of continuum resonances and extraction of the instantaneous emission rate. a,** Probability density after projecting out the ground state (same as Extended Data Fig. 7b). **b-d,** Remaining density after successively removing the respective resonances (shown on the right). **e,** yield (orange) and instantaneous rate (blue) evaluated after removing $\Psi_{res}^{(0)}$ to $\Psi_{res}^{(3)}$ as shown in panel (d).

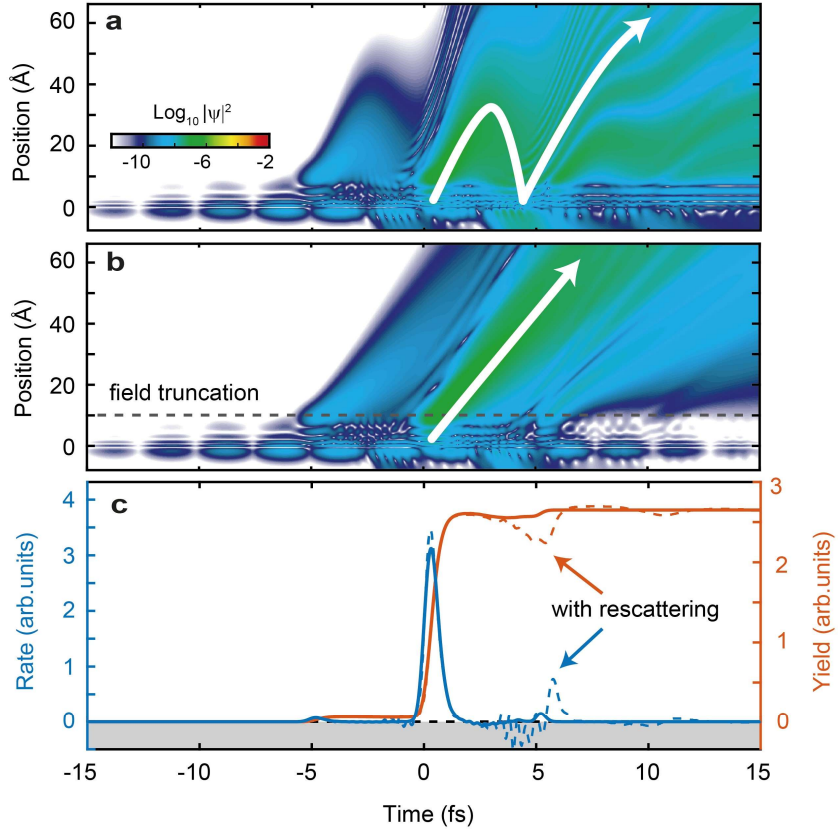

**Extended Data Figure 9| Comparison between full and truncated-field time propagation. a,b**, Probability density of the wavefunction after removing $\Psi_{res}^{(0)}$ to $\Psi_{res}^{(3)}$ including (a) and without (b) rescattering by truncating the field for positions $x > 10$ Å (dashed line). **c,** Yield (orange) and rate (blue) determined from (a) (dashed lines) and (b) (solid lines).

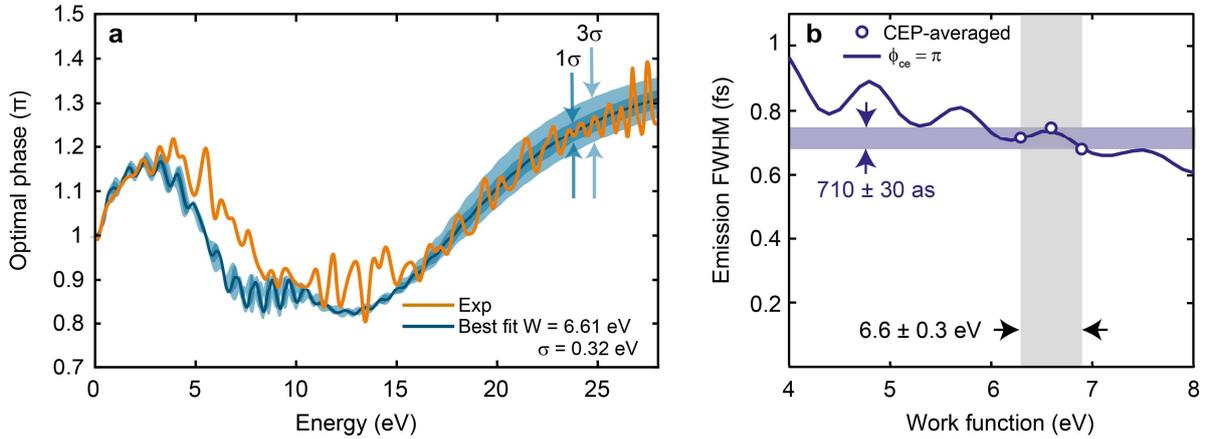

**Extended Data Figure 10| Least square optimization of work function and mapping to emission duration. a,** The minimum in the least squares optimization is found at $W = 6.61$ eV (blue curve) and matches the experimental optimal phase (orange) in the cut-off domain. The $1\sigma$ and $3\sigma$ confidence intervals are indicated for $\sigma = 0.32$ eV. **b,** Emission duration (FWHM) depending on work function for $\phi_{ce} = \pi$. Blue circles indicate CEP-averaged durations for $W = 6.3$ eV, 6.6 eV and 6.9 eV. The grey area indicates the previously determined work function interval. The maximum variation of the emission duration within this interval of $710 \pm 30$ as is highlighted by the blue shaded area.

# Supplementary Information: Tracing attosecond electron emission from a nanometric metal tip

Philip Dienstbier[1], Lennart Seiffert[2,*], Timo Paschen[1,3,*], Andreas Liehl[4], Alfred Leitenstorfer[4], Thomas Fennel[2,5,6], and Peter Hommelhoff[1]

[1]*Department of Physics, Friedrich-Alexander-Universität Erlangen-Nürnberg (FAU), Staudtstraße. 1, D-91058 Erlangen, Germany, EU*

[2]*Institute for Physics, Rostock University, Albert-Einstein-Straße 23–24, D-18059 Rostock, Germany, EU*

[3]*Now with: Korrelative Mikroskopie und Materialdaten, Fraunhofer-Institut für Keramische Technologien und Systeme IKTS, Äußere Nürnberger Straße 62, D-91301 Forchheim, Germany, EU*

[4]*Department of Physics and Center for Applied Photonics, University of Konstanz, D-78457 Konstanz, Germany, EU*

[5]*Max Born Institute, Max-Born-Straße 2A, D-12489 Berlin, Germany, EU*

[6]*Department of Life, Light and Matter, University of Rostock, Albert-Einstein-Straße 25, D-18059 Rostock, Germany, EU*

**Emission duration and time-energy uncertainty relation**

The emission duration of 710 as agrees with the time-energy uncertainty relation, which we demonstrate in the following. Figure S1a shows the photoelectron spectrum obtained by the TDSE if rescattering is suppressed. Due to the field truncation a net ponderomotive energy remains and shifts the complete spectrum[1]. Converting the energy axis to frequency using $E = hf$ results in the spectrum shown in Fig. S1b. In analogy to the Fourier-limit of optical pulses we Fourier transform the square root of the spectrum assuming a flat spectral phase. The square modulus of the Fourier transform is shown in Fig. S1c and is referred to as the Fourier spectrum. The Fourier spectrum consist of three peaks spaced by the optical cycle duration of 5.2 fs following from the ATI peak structure with a spacing of 0.8 eV in Fig. S1a. The FWHM of the inner peak amounts to 600 as and thus is slightly shorter than the result from our rate extraction method.

The full photoelectron spectrum in the TDSE and experiment in Fig. S1d contains energy features from both, the directly emitted electrons but also rescattered electrons. However, only the energy distribution of directly emitted electrons is directly related to the emission duration. Hence, we consider energies up to $2U_\mathrm{p}$ and mirror the spectrum in the negative frequency domain as shown in Fig. S1e. Mirroring the spectrum is required as only half of the initially launched wavepacket forms the direct part of the spectrum, whereas the other half of initially present momenta/energies are redistributed by rescattering. The resulting transform limited durations of 660 as from the TDSE and 560 as from the experimental spectrum shown in Fig. S1f are shorter than the result from the rate extraction method. To conclude, the results of the rate extraction is valid within the time-energy uncertainty relation and is close to the transform limit.

---

[*]Both authors contributed equally.

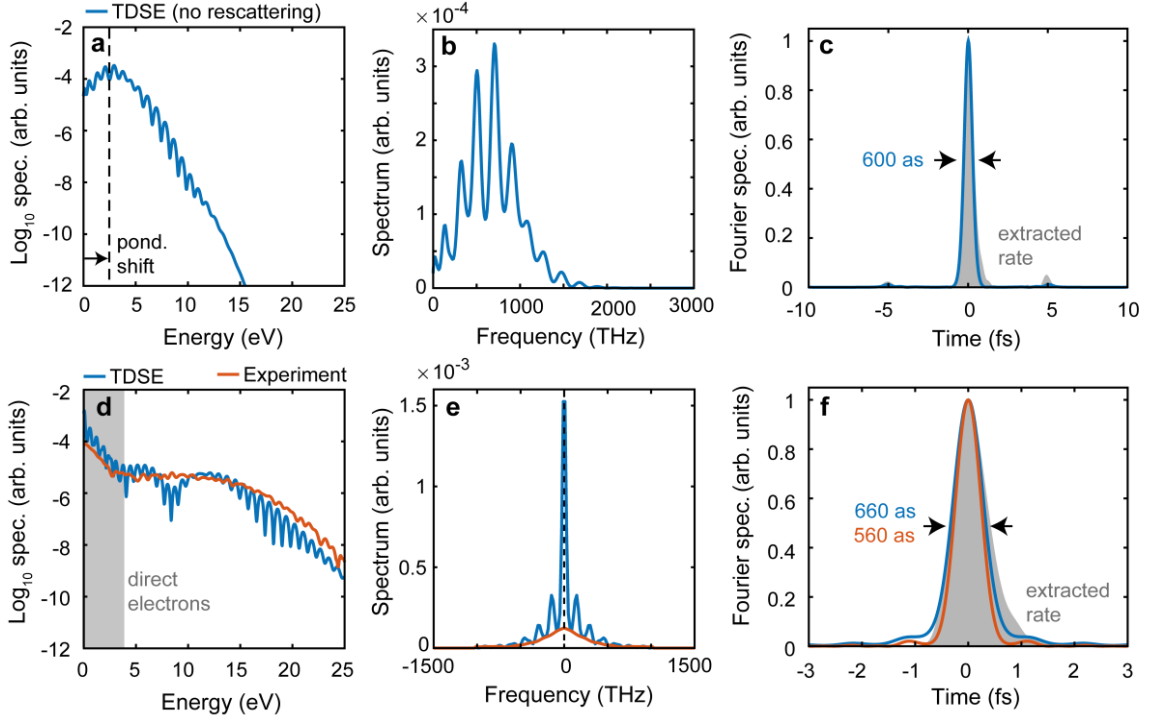

**Figure S1| Emission duration and time-energy uncertainty relation. a,** Simulated photoelectron spectrum with suppressed backscattering. The truncated potential (Extended Data Fig. 9) causes an average momentum/energy of the leaving wave packet. **b,** Spectrum from **a** in linear scale and energy axis converted to frequency. **c,** Fourier spectrum (blue) assuming a flat spectral phase resulting in a Fourier-limited duration of 600 as being slightly shorter than the result from the rate extraction (grey). **d,** Photoelectron spectrum for $\phi_{\text{rel}} = \pi$ and $\phi_{\text{ce}} = \pi$ from the TDSE and experimental spectrum for $\phi_{\text{rel}} = \pi$ and averaged CEP. **e,** Spectra from **d** in linear units with energy axis converted again to frequency. The negative frequencies are mirrored and only spectral components below an energy of $2U_{\text{p}}$ have been considered. **f,** Fourier spectrum assuming a flat spectral phase resulting in a Fourier limited durations of 660 as for the TDSE and 560 as for the experiment.

**Laser induced work function change**

The retrieved work function of $W = (6.6 \pm 0.3)$ eV significantly deviates from the literature value of 4.35 eV for a clean tungsten (310) surface[2]. In the following, we show that the work function rises after minutes of illumination with strong few-cycle pulses with a high repetition rate of 100 MHz, despite the ultrahigh vacuum environment. To independently measure the work function, we illuminate the tip with two continuous wave fields at the wavelengths of $\lambda_1 = 405$ nm and $\lambda_2 = 488$ nm. By increasing the tip bias, the potential barrier is lowered until the onset of single-photon emission is observed as shown in Fig. S11. The effective barrier height is given by

$$W_{\text{eff}} = W - \sqrt{\frac{e^3 |U_{\text{tip}}|}{4\pi\epsilon_0 k r}} \quad (1)$$

involving the work function $W$ and the field reduction factor $k$. The field reduction factor $k$ describes the deviation of the local static field $E_{DC} = \frac{U_{tip}}{kr}$ from a spherical geometry[3]. The energy difference between both wavelengths $\Delta E = hc\left(\frac{1}{\lambda_1} - \frac{1}{\lambda_2}\right)$ needs to match the difference of the effective work function for the onset voltages $U_{tip,1}$ and $U_{tip,2}$, which allows us to calculate

$$k = \frac{\left(U_{tip,1} + U_{tip,2} - 2\sqrt{U_{tip,1} U_{tip,2}}\right)e^3}{4\pi\epsilon_0 r \, \Delta E^2} \text{ and } W = \frac{hc}{\lambda_1} + \sqrt{\frac{e^3|U_{tip,1}|}{4\pi\epsilon_0 kr}}. \quad (2)$$

We apply this method to two cases: (1) right after cleaning to an atomically clean surface, and (2) after the tip was illuminated like in the experiment. Fig. S2a shows the work function and field reduction factor are obtained before the tip is illuminated with the strong few-cycle pulses from the fibre laser. Clearly, they match the expectation for a clean (310) facet (W = 4.3 eV) and standard tip geometry[4] with $k \sim 3 - 8$. Repeating the measurement after about 20 minutes of illumination with the few-cycle pulses yields the results shown in Fig. S2b: We observe a rather drastic change of the onset voltages. The corresponding work function is $W = (6.4 \pm 0.6)$ eV, in excellent agreement with the result from the optimal phase in the two-colour measurement. The field reduction factor also changes towards a more hemispherical emission geometry. When we apply a large positive bias voltage (~7.5kV) the surface is cleaned again by field evaporation and, indeed, the work function can be changed back to its initial value. The work function change is accompanied by a smeared-out emission pattern and reduced emission yield.

Large work functions of 6.2 eV and 6.0 eV after illuminating the tip with strong laser pulses at high repetition rates have been reported[5,6] and were attributed to adsorbates on the tip. However, also oxides or distortions of the last few lattice layers may change the work function. The microscopic root cause of the work function change needs to remain for future work. While we do not know the exact nature of work function increase, we consider it unlikely that the microscopic surface variation will lead to resonant tunnelling behaviour[7].

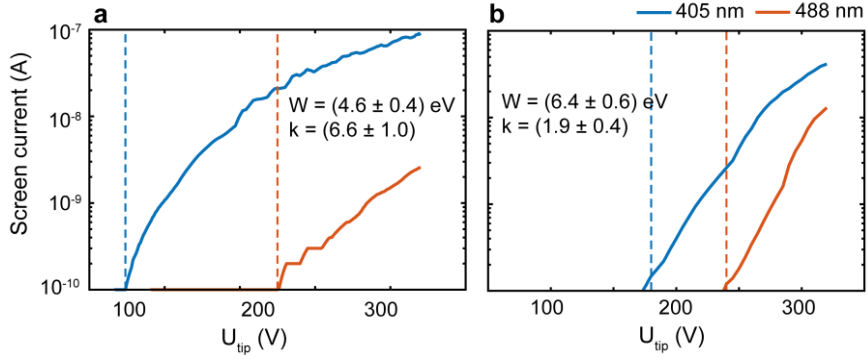

**Figure S2| Independent work function determination with the help of photoemission experiments. a,** Emission current as a function of tip bias voltage $U_\text{tip}$ for a wavelength of 405 nm (blue) and 488 nm (orange) right after in-site cleaning with field evaporation and before the tip was illuminated with strong few-cycle pulses. The onset of single-photon emission for the two wavelengths (dashed lines) indicates a work function of $W = (4.6 \pm 0.4)$ eV. We further obtain a field reduction factor $k = 6.6 \pm 1.0$. **b,** After illuminating the tip with few-cycle pulses for about 20 minutes, the barrier-lowering required for single-photon emission changes notably. We now obtain a work function of $W = (6.4 \pm 0.6)$ eV and field reduction factor $k = 1.9 \pm 0.4$. Hence, these independent measurements fully support the work function of $(6.6 \pm 0.3)$ eV of the main text.

**Robustness of the optimal phase analysis**

In the main text, we showed that the characteristic energy dependence of the optimal phase within the direct part, the plateau, and the cut-off domain can be understood by combining final energies of classical trajectories with instantaneous rates obtained from quantum simulations (Fig. 3 and Extended Data Fig. 3). In this section, we demonstrate the robustness of the optimal phase analysis by investigating a set of additional simulations for different wavelengths and intensities.

As a reference, we replot the simulated optimal phases for the parameters considered in the main text (cf. Fig. 3c) in Fig. S3a and the corresponding work function-dependent emission durations in Fig. S3b (coloured symbols, $\phi_\text{ce} = \pi$) including the symmetrized error boundary discussed in Methods. Figures S3c-f show respective results for simulation runs for doubled intensities (panels c,d) and increased wavelengths (panels e,f with $\lambda_\omega = 2000$ nm and $\lambda_{2\omega} = 1000$ nm).

For all considered cases, the optimal phase exhibits the same characteristic evolution: above $\pi$ in the energy domain of direct emission $(0 - 2\ U_\text{p})$, below $\pi$ in the plateau $(2U_\text{p} - 10\ U_\text{p})$, and above $\pi$ in the cut-off domain $(> 10\ U_\text{p})$. In the latter, the optimal phase is most sensitive to the work function, which allows the precise matching of the theory model to the experiment and enables to extract the emission duration.

While our analysis shows that the optimal phase and our method are robust in the considered range, the extracted emission durations may vary depending on the exact parameters. For example, at the larger intensity the emission duration is overall shorter, while for larger wavelengths it is increased, which in both cases is expected from an analytical emission model[8].

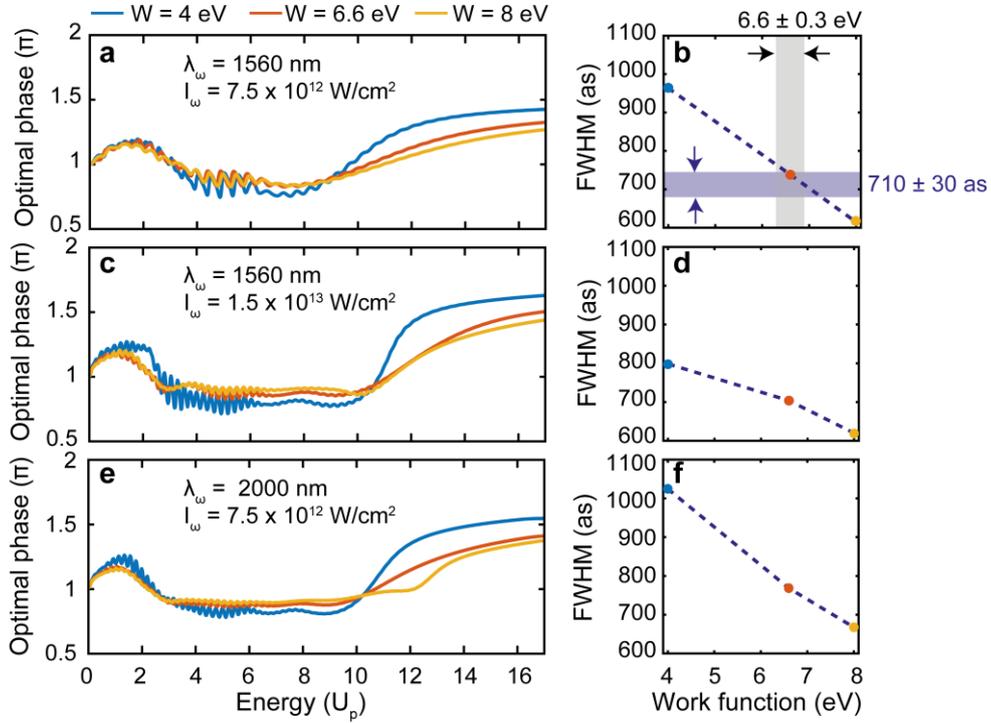

**Figure S3| Robustness of the optimal phase analysis. a,** Energy-dependent optimal phases extracted from TDSE simulations for parameters as in the main text (cf. Fig. 3c) for three different work functions as indicated. The electron energy is given in units of the ponderomotive energy $U_\text{p}$. **b,** Corresponding emission durations for $\phi_\text{ce} = \pi$. Shaded areas show the error boundaries for the work function and corresponding CEP-averaged emission duration as in Extended Data Fig. 10b. **c, d,** Same as a and b, but for twice the fundamental intensity ($I_\omega = 1.5 \times 10^{13}\,\text{W/cm}^2$, same admixture). **e, f,** Same as a and b, but for longer fundamental and second harmonic wavelengths of $\lambda_\omega = 2000$ nm and $\lambda_{2\omega} = 1000$ nm (same admixture). Dashed lines in the right panels serve as guides to the eye.

**Optimal phase analysis using the carrier-envelope phase**

The key result of our study is the extraction of the emission duration via the characteristic optimal phase, which originates from a selective modulation of energies and yields in measured photoelectron spectra by the relative phase of a two-colour field. Instead of this relative phase of our two-colour experiment, the carrier-envelope phase of a single-colour few-cycle driving field could provide an alternative approach to imprint similar signatures into electron spectra. To find out if this approach is also feasible, we investigate simulated CEP-dependent electron spectra as shown in Fig. S4a-c for three work functions as indicated. The spectra show prominent CEP-dependent signatures motivating the definition of an energy-dependent optimal CEP $\phi_{ce}^{opt}$ (black curves in a-c), in analogy to the optimal relative phase in the two-color scheme. Comparison of the optimal CEPs for the three work functions in Fig. 4d shows that the optimal CEPs stay below $\pi$ in the direct emission domain ($<2U_p$) followed by an overall growth in the plateau, and saturation in the cut-off domain at around $4\pi$. While this trend is similar for all work functions, the final saturation value differs depending on the work function (see inset in Fig. 4d) with an optimal CEP variation of $\Delta\phi_{ce}^{opt} \approx 0.15\pi$ between the work functions $W = 4$ eV and $W = 8$ eV. This optimal CEP variation is comparable to the two-colour result and should therefore also allow extracting the work function by matching simulation results to experimental data.

In the following, we show that also this variation of the optimal CEP can be related to the emission duration by performing the combined trajectory and rate analysis in Fig. S5 from the two-colour study (cf. Fig. 3). Figures S5a,d show the final energies of classical electron trajectories depending on birth time and CEP, while Figs. S5b,e show corresponding rates determined from TDSE simulations for two work functions as indicated. The separated diagonal features correspond to electrons born in different cycles of the few-cycle field. The maximum final energy (red square) is generated in the half-cycle preceding the peak envelope as it rescatters during the central cycle. However, the highest rate is reached in the central cycle.

To clarify the physical origin of the optimal CEP in the cut-off region, we select trajectories with final energies above 9.6 $U_p$ and integrate their corresponding rates along the birth time axis (cf. Fig. 3) resulting in the CEP-dependent yields shown in Figs. 5c,f. The optimal CEPs defined via the location of the maximum yields are indicated as grey circles and differ from the phase realizing the maximum energy. Our analysis shows that the instantaneous rate is stronger localized to the inner cycle for the higher work function. Hence, the overlap between birth times and CEPs leading to final energies in the cut-off domain and large rates shifts towards the inner cycle (arrow in Fig. 5e) resulting in a smaller optimal CEP. This established the link between the variation of the optimal CEP and the rate similar to the two-colour scheme, which should enable the extraction of the emission duration also using the CEP of few-cycle fields.

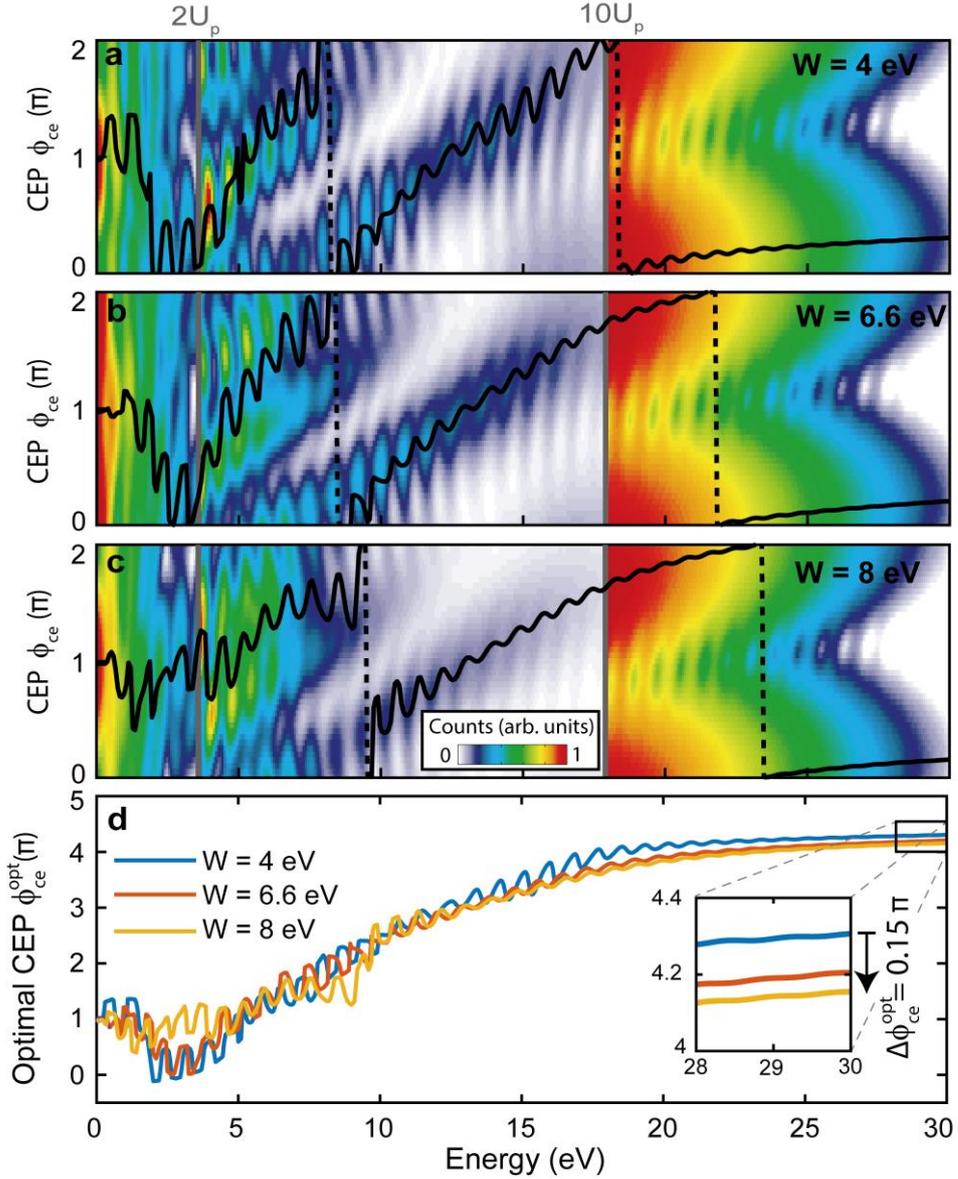

**Figure S4| Optimal carrier-envelope phase analysis in a few-cycle field. a-c,** Simulated electron energy spectra depending on the carrier-envelope phase $\phi_{ce}$ of a few-cycle pulse (same parameters as fundamental field in the two-color scheme, see Methods) for three work functions, as indicated. The same visualization as in Fig. 2 of the main text is used. Black curves show energy-dependent optimal carrier-envelope phases $\phi_{ce}^{opt}(E)$ determined by harmonic fits (cf. Extended Data Fig. 1). Vertical dashed lines indicate jumps by $2\pi$. **d,** Unwrapped optimal CEPs $\phi_{ce}^{opt}$ from panels a-c. The inset indicates a shift of $\Delta\phi_{ce}^{opt} \approx 0.15\,\pi$ towards lower optimal CEPs when increasing the work function from 4 to 8 eV.

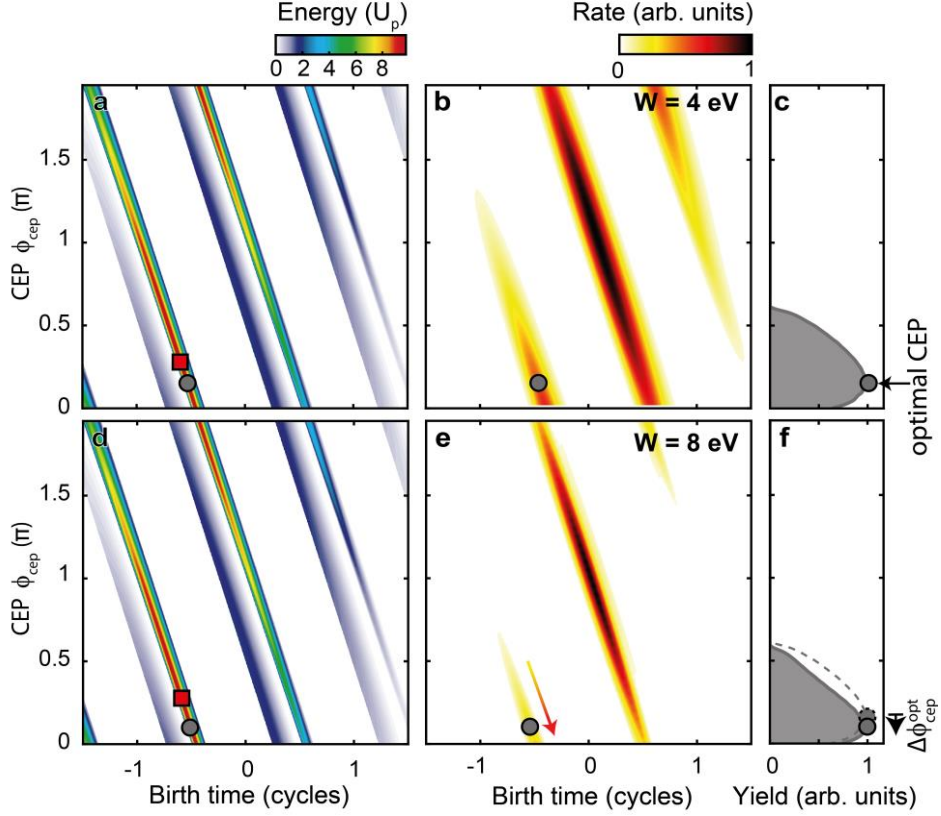

**Figure S5| Origin of the optimal CEP. a,d,** Final energies of electron trajectories as a function of the carrier-envelope phase and birth time obtained using the simple man's model (in analogy to Fig. 3b). The red square indicates the birth time and CEP when the maximum final energy is reached. **b,e,** Instantaneous rates for the same birth time and CEP window obtained from the TDSE (see Methods) for two work functions, as indicated. **c,f,** CEP-dependent yield obtained by integrating along the birth time axis in b,e considering final energies $\geq 9.6 U_p$ deduced from a,d. The optimal carrier-envelope phases are indicated by grey circles. For comparison, the results from panel c are inserted in panel f as dashed curve and symbol.

**Simulations based on the time-dependent Schrödinger equation**

For the quantum simulations we use the model described in Ref.[10] (Ref.[27] in the manuscript, also used in minimal code for the rate extraction[11]), where we consider the one-dimensional time-dependent Schrödinger equation (TDSE) in single active electron approximation and length gauge. The wavefunction $\Psi(x,t)$ is obtained by integrating the TDSE

$$i\hbar \frac{\partial}{\partial t}\Psi(x,t) = \left[-\frac{\hbar^2}{2m}\frac{\partial}{\partial x^2} + V(x,t)\right]\Psi(x,t) \quad (1)$$

using the Crank-Nicolson method with the electron mass $m$, reduced Planck constant $\hbar$ and potential $V(x,t)$. The potential

$$V(x,t) = V_0(x) + V_i(x,t) = V_0(x) + \int_0^x E_{\omega-2\omega}(x',t)dx' \quad (2)$$

includes the binding potential $V_0(x)$ and the light-matter interaction potential $V_i(x,t)$. For $V_0(x)$ we consider a box potential with depth $\tilde{V}_0 = W + E_F$, where the width is chosen to match a desired work function $W$ for a fixed Fermi-energy of $E_F = 7$ eV. As the initial state we chose the ground state $\Psi_{\text{ground}}(x)$ determined via imaginary time propagation, which for the chosen parameters is the only bound state. The light matter-interaction potential includes the spatially dependent near-fields

$$\gamma_{\omega/2\omega}(x) = \begin{cases} 0 & \text{for } x < 0 \\ 1 + (\gamma^0_{\omega/2\omega} - 1)\exp(-x/\lambda_{\text{nf}}) & \text{for } x \geq 0 \end{cases} \quad (3)$$

with common decay length $\lambda_{\text{nf}} = 100$ Å for both colours[9] and field enhancement factors $\gamma^0_\omega$ and $\gamma^0_{2\omega}$ as determined from the experiment. The two-colour near-field is defined as

$$\begin{aligned} E_{\omega-2\omega}(x,t) &= E^{\text{inc}}_\omega \gamma_\omega(x) f_\omega(\omega t) \cos(\omega t + \phi_{\text{ce}}) \\ &+ E^{\text{inc}}_{2\omega} \gamma_{2\omega}(x) f_{2\omega}(2\omega t + \phi_{\text{rel}}) \cos(2\omega t + 2\phi_{\text{rel}} + \phi_{\text{ce}}) \end{aligned} \quad (4)$$

with the relative phase $\phi_{\text{rel}}$ and carrier-envelope phase $\phi_{\text{ce}}$. The envelopes are defined as

$$f_{\omega/2\omega}(\phi) = \exp\left(-2\ln 2 \left(\frac{\phi}{\tau^{\text{cycle}}_{\omega/2\omega}}\right)^2\right) \quad (5)$$

where the pulse durations $\tau_{\omega/2\omega}$ are related to their durations in cycles $\tau^{\text{cycle}}_\omega = \omega \tau_\omega$ and $\tau^{\text{cycle}}_{2\omega} = (2\omega)\tau_{2\omega}$. Absorbing boundary conditions are used on both sides of the simulation domain to avoid reflections. Electron spectra are calculated at the end of the simulation via the window operator method and are CEP-averaged as the laser system is not CEP-stable.

We note that including the relative phase in both the envelope and the carrier of the second harmonic is equivalent to a temporal delay which has been considered for all calculations within this work. However, hardly any changes of the simulated spectra within the experimentally considered delay range are notable when only including the phase in the carrier. This signifies that the relative phase is responsible for the modulation of the spectra while modifications due to the envelope offset are negligible. This is mainly due to the near three-cycle duration of the second harmonic field and CEP-averaging.

We further note that hardly any changes in the electron spectra and optimal phases are present if a spatially homogenous two colour near-field is assumed, which simplifies the light-matter interaction potential to

$$V_i(x,t) = \begin{cases} 0 & \text{for } x < 0 \\ x E_{\omega-2\omega}(x=0,t) & \text{for } x \geq 0 \end{cases} \quad (6)$$

involving the field directly at the metal-vacuum interface

$$\begin{aligned}E_{\omega-2\omega}(x=0,t) &= E_\omega(f_\omega(\omega t)\cos(\omega t+\phi_{\text{ce}}) \\ &\quad + \alpha f_{2\omega}(2\omega t+\phi_{\text{rel}})\cos(2\omega t+\phi_{\text{rel}}+2\phi_{\text{ce}})),\end{aligned} \quad (7)$$

which only depends on the peak fundamental near-field $E_\omega$ and admixture $\alpha$ as used in the main text. The negligible impact of the spatial field profile justifies the use of homogeneous near-fields within the trajectory model and within the rate extraction described in methods.